\documentclass{jpp}
\usepackage{graphicx}

\usepackage[utf8]{inputenc}
\usepackage[T1]{fontenc}
\usepackage{amsmath}
\usepackage{xcolor}

\def \be {\mathbf{e}}
\def \bk {\mathbf{k}}
\def \bv {\mathbf{v}}
\def \bx {\mathbf{x}}
\def \bB {\mathbf{B}}

\graphicspath{{figures_13_jpg/}}

\shorttitle{Effect of self-interaction on zonal flows}
\shortauthor{Ajay C. J., S. Brunner, B. McMillan, J. Ball, J. Dominski, G. Merlo}

\title{How eigenmode self-interaction affects zonal flows and convergence of tokamak core turbulence with toroidal system size}

\author{Ajay C.J.\aff{1}
  \corresp{\email{ajay.chandrarajanjayalekshmi@epfl.ch}},
  Stephan Brunner\aff{1},
  Ben McMillan\aff{2},
  Justin Ball\aff{1},
  Julien Dominski\aff{3},
 \and Gabriele Merlo\aff{4}}

\affiliation{\aff{1}Ecole Polytechnique F\'ed\'erale de Lausanne (EPFL), Swiss Plasma
  Center, CH-1015 Lausanne, Switzerland
\aff{2}Centre for Fusion, Space and Astrophysics, Department of Physics,
  University of Warwick, CV4 7AL, Coventry, UK
\aff{3}Princeton Plasma Physics Laboratory, Princeton University, P.O. Box
  451, Princeton, New Jersey 08543-0451, USA
\aff{4}The University of Texas at Austin, Austin, Texas 78712, USA}

\begin{document}

\maketitle

\begin{abstract}

Self-interaction is the process by which a microinstability eigenmode
that is extended along the direction parallel to the magnetic field
interacts non-linearly with itself. This effect is particularly
significant in gyrokinetic simulations accounting for kinetic passing
electron dynamics and is known to generate {\it stationary} $E\times
B$ zonal flow shear layers at radial locations near low order mode
rational surfaces \citep{Weikl2018}. We find that self-interaction in
fact plays a very significant role in also generating {\it
  fluctuating} zonal flows, which is critical to regulating turbulent
transport {\it throughout} the radial extent. Unlike the usual picture
of zonal flow drive in which microinstability eigenmodes coherently
amplify the flow via modulational instabilities, the self-interaction
drive of zonal flows from these eigenmodes are uncorrelated with each
other. It is shown that the associated shearing rate of the
fluctuating zonal flows therefore reduces as more toroidal modes are
resolved in the simulation. In 
simulations accounting for the full toroidal domain, such an increase
in the density of toroidal modes corresponds to an increase in the
toroidal system size, leading to a finite system size effect that is distinct
from the well-known profile shearing effect.

\end{abstract}

\section{Introduction}

Microturbulence driven by small Larmor-scale instabilities is in most
cases the dominant cause of heat and particle loss from the core of
magnetic confinement fusion devices and therefore presents a major
challenge in achieving burning plasma conditions
\citep{Horton1999}. The role of passing electron dynamics in turbulent
transport driven by ion-scale microinstabilities, in particular Ion
Temperature Gradient (ITG) and Trapped Electron Mode (TEM)
instabilities, has been given relatively little attention. In first
approximation, these particles, which are highly mobile along the
confining magnetic field, are assumed to respond adiabatically to the
low frequency ion-scale modes. However, in a tokamak geometry, the
phase velocity parallel to the magnetic field of a perturbation with
fixed mode numbers $m$ and $n$ in the poloidal and toroidal
directions, respectively, becomes infinite at the radial position
$r_{m,n}$ of the corresponding Mode Rational Surface (MRS), where the
magnetic safety factor profile $q_s(r)$ is such that $q_s(r_{m,n}) =
m/n$. Consequently, at these positions, the condition for passing
electrons to respond adiabatically gets violated and their
non-adiabatic response becomes important. One notes that, given the
toroidal axisymmetry of a tokamak, each linear microturbulence
eigenmode has a fixed toroidal mode number $n$, but is in general a
superposition of many poloidal Fourier modes $m$, so that associated
MRSs are the positions $r_{m,n}$ for the different values of $m$.

In fact, as a result of the non-adiabatic passing electron dynamics,
linear ITG and TEM eigenmodes can become significantly extended along
the magnetic field lines \citep{Hallatscheck2005}, producing fine
radial structures at corresponding MRSs
\citep{Falchetto2003,Chowdhury2008}. These fine structures on the
eigenmodes persist in corresponding turbulence simulations, and via
non-linear couplings lead to stationary corrugations on the radial
profiles of density, temperature and in particular $E\times B$ zonal
shear flows, aligned with low order MRSs, which effectively appear as
fine-scale transport barriers
\citep{Waltz2006,Dominski2015,Dominski2017}. Given that shearing and
decorrelation of turbulent eddies by zonal flows is a primary
mechanism by which turbulence saturates
\citep{Biglari1990,Waltz1994,Rosenbluth1998,Lin1998}, it is clearly of
interest to understand the generation of these fine scale zonal flow
structures in detail.
         
The generation of the fine zonal flow structures at low order MRSs, via a process called self-interaction, has in fact already been discussed to some length in the work by \citet{Weikl2018}. The general motivation for the work
presented in this paper has been to further investigate the properties
of zonal flow drive from this self-interaction mechanism. This process essentially
involves each individual microturbulence
eigenmode interacting non-linearly with itself to produce a Reynolds
stress contribution to the zonal flow drive, which turns out to be
located around its corresponding MRSs. Key to this self-interaction is
the fact that an eigenmode which is significantly extended along the
magnetic field line `bites its tail' after one poloidal turn. For each
eigenmode, self-interaction will therefore be particularly significant
at its MRSs, given that the radially fine structures located at these
positions, resulting from the non-adiabatic passing electron response,
are elongated along the magnetic field lines. At low order MRSs (in
practice implying mainly integer and half-integer $q_s$-surfaces),
these self-interaction contributions to Reynolds stress from the
different eigenmodes radially align and add up constructively to drive
the stationary $E\times B$ zonal shear flows.
%% (note in particular that lowest order MRSs are common to all
%% eigenmodes).
Clear illustrations of these fine stationary zonal flow structures can
be seen for example in figure~2 of Ref.  \citep{Waltz2006}, figure~12
of Ref. \citep{Dominski2017}, figure~3 of Ref.  \citep{Weikl2018}, and
figure~\ref{recap} of this paper. 

It is important to point out that in the core of a tokamak, low order
MRSs are in fact few and far apart, whereas the radial domains between
them occupy the majority of the plasma volume. Therefore, at least as
important as understanding how self-interaction generates the
above-mentioned stationary structures at lowest order MRSs, clearly
visible on any flux surface-averaged field, is to address how the same
self-interaction mechanism might affect zonal flow drive and therefore
turbulent transport in the radial regions {\it between} low order
MRSs. Addressing this question is the main focus of this paper.

While lowest order MRSs are common to all eigenmodes, between them,
the MRSs related to the various microturbulence modes tend to be
radially misaligned. Hence, in these radial regions, the
self-interaction contributions to Reynolds stress from these modes
tend to cancel each other out when averaged over time, thus resulting
in a nearly \emph{zero stationary} component of the zonal flows. But,
given that in the turbulent phase amplitudes of the various
microinstability eigenmodes vary in time and furthermore are
decorrelated with each other, this cancellation is not ensured at each
time instant but only on average over time. As a result,
self-interaction drives a \emph{non-zero fluctuating} zonal flow
component between low order MRSs.
%% (as opposed to the zero \emph{stationary} component at these
%% radial positions).

An important observation resulting from our study is that in
simulations accounting for non-adiabatic passing electron dynamics,
self-interaction contributions to Reynolds stress from the various
microturbulence modes can indeed be significant in the radial regions
between lowest order MRSs and in fact act as \emph{random decorrelated
  kicks} that can in some cases actually disrupt what is usually
considered the main drive mechanism of zonal flows - modulational
instability \citep{Hasegawa1978,Hasegawa1979,Chen2000}. The
modulational instability mechanism involves the resonant decays of
microinstability modes into zonal modes via secondary microinstability
daughter modes. Unlike self-interaction, it is a \emph{coherent}
process, leading to a {\it correlated} contribution from the various
microinstability modes to the Reynolds stress drive of zonal modes.

In order to quantify the relative importance of the two alternative,
possibly competing mechanisms driving zonal flows, {\it i.e.}
self-interaction and modulational instability, we have studied the
statistical properties of the Reynolds stress contributions from the
various microturbulence modes. Given the different nature of the two
driving processes, one can identify high correlation levels between
the mode contributions to conditions where modulational instability
drive dominates, while lower correlation levels are characteristic of
a significant effect from self-interaction. Central to this
statistical analysis has been a series of gyrokinetic simulations
obtained for identical driving conditions but varying the number of
significant toroidal modes participating in the turbulence. Varying
the number of toroidal modes changes in particular the number of
associated random kicks from self-interaction to the zonal flow drive
at each radial position.
%% Given the different nature of the zonal flow drive from the two
%% alternative, possibly competing mechanisms, {\it i.e.}
%% self-interaction and modulational instability, we have studied the
%% statistical properties of the Reynolds stress contributions from the
%% various microturbulence modes to quantify the relative importance of
%% these two processes; conditions where the drive from modulational
%% instabilities dominates are characterized by high levels of
%% correlation between these contributions, while conditions dominated by
%% drive from self-interaction lead to low correlations.
%%
%% In order to quantify the relative importance of self-interaction and
%% modulational instability as two alternative, possibly competing
%% processes driving zonal flows, and given their different nature (the
%% first providing decorrelated contributions to the drive, the latter
%% correlated ones), we study the statistical properties of the Reynolds
%% stress contributions from the various microturbulence modes.
%% To this end, we vary the number of significant toroidal modes
%% participating in the turbulence simulations.

Varying the number of significant toroidal modes is in fact related to
varying the system size, typically measured by $\rho^\star =
\rho_i/a$, {\it i.e.} the ratio of the thermal ion Larmor radius
$\rho_i$ to the minor radius $a$ of the tokamak. Indeed, invoking the
fact that the unstable modes driving ion-scale microturbulence are
such that $k_\perp\rho_i \lesssim 1$, where $k_\perp$ is the wave
vector component perpendicular to the magnetic field, and furthermore
noticing that for a given eigenmode with toroidal mode number $n$ an
estimate for $k_\perp$ is given by the poloidal wave vector component,
which in turn can be evaluated as $k_\chi\simeq m/r_0\simeq
nq_s(r_0)/r_0$ [$r_0$ is the average radial position of the eigenmode
  and $m$ its characteristic poloidal mode number estimated as
  $m\simeq nq_s(r_0)$, given that fluctuations are nearly
  field-aligned], one obtains that $k_\perp\rho_i \simeq
nq_s(r_0)\rho_i/r_0 = (q_s a/r_0) n\rho^\star \sim n\rho^\star $, so
that the number $N_\varphi$ of toroidal modes contributing to the
turbulence scales as $N_\varphi\sim 1/\rho^\star$.

In view of what has just been said, it may at first sight appear
surprising that the study presented in this paper is in fact based on
simulations carried out in the framework of the flux-tube model
\citep{Beer1995}, whose underlying assumption is the scale separation
between the characteristic length of microturbulence ($\sim$ Larmor
radius $\rho$) and the characteristic length of equilibrium ($\sim$
minor radius $a$), hence {\it a priori} achieved by taking the strict
limit $\rho^\star\to 0$ of the gyrokinetic equations. In this limit,
the radial dependence of all equilibrium profiles and their gradients
appearing in the gyrokinetic equations are constant over the flux-tube
volume. There is however one exception to this in the standard flux
tube model \citep{Scott1998}: a linearized radial dependence of the
safety factor profile $q_s(r)$ ($\sim$ constant magnetic shear
approximation) is kept in the twist and shift boundary condition
applied after usually following the magnetic field line a single turn
in the poloidal direction. Through this parallel boundary condition,
an eigenmode 'bites its tail' and in particular correctly accounts for
the possible resonances that may develop at its associated MRSs. The
fact that the eigenmodes correctly ``feel'' their associated MRSs is
obviously key to the development of the fine radial structures and
related non-linear self-interaction which is the focus of our
study. Also, close agreement between flux-tube \citep{Dominski2015}
and global \citep{Dominski2017} gyrokinetic simulations regarding the
characteristics of the fine stationary zonal flow structures at lowest
order MRSs clearly confirm that the dynamics of the self-interaction
mechanism can indeed be correctly accounted for using such a local
model.

Let us already briefly provide here some further insight into how the
toroidal spectral density is varied in a flux-tube code (all details
will be given in \S \ref{secGENE}). 
The flux-tube version of the GENE code \citep{GENE1,GENE3}
which we have applied for the present study, considers the field aligned coordinate system defined [in equations~(\ref{def. x coordinate}-\ref{def. z coordinate})] by the variables $(x, y, z)$ which are respectively the
radial, binormal and parallel coordinates.
%% A magnetic field line is defined by $x,y=$ constant, in other words
%% the equilibrium magnetic field $\vec{B}_0$ is parallel to $\nabla
%% x\times \nabla y$. The flux-tube simulation volume is then simply a
%% box of size $L_x\times L_y \times L_z$ in $(x, y, z)$ coordinates,
%% which may be visualized in configuration space as a twisting
%% field-aligned plasma sub-volume lying on the magnetic surface $r=r0$,
%% sheared as a result of finite magnetic shear $\hat{s}$, and usually
%% carrying out one poloidal turn if $L_z=2\pi$.
Note that the binormal coordinate $y$ is the only one of the three
coordinates $(x, y, z)$ which actually depends on the toroidal angle
$\varphi$. The length $L_y$ of the flux-tube in this direction can
thus be associated to its angular extent $\Delta\varphi$ along
$\varphi$: $L_y = C_y\Delta\varphi$, where the constant $C_y = r_0/q_0$ ensures that $y$ has units of length, $r_0$ being the radial position at which the flux-tube is positioned and $q_0 = q_s(r_0)$. Also, as a linear eigenmode has a
fixed toroidal mode number $n$, it consequently has a fixed Fourier
mode number $k_y$ with respect to $y$ when represented in
field-aligned coordinates, given by $k_y = n/C_y = nq_0/r_0$, {\it
  i.e.}  $k_y$ is actually equivalent to the previously mentioned
poloidal mode number estimate $k_\chi$. The significant $k_y$ modes
contributing to ion-scale turbulence are therefore such that $k_y\rho_i
\lesssim 1$, and given that the minimum mode number for a flux-tube
box with length $L_y$ is given by $k_{y,\min}=2\upi/L_y$, the number
of toroidal modes contributing to the turbulence is estimated as
$N_\varphi\simeq 1/k_{y,\min}\rho_i= (1/2\upi)L_y/\rho_i$. Hence, when
studying microturbulence using flux-tube simulations, carrying out a
scan in $k_{y,\min}\rho_i$ (or equivalently $L_y/\rho_i$) corresponds
to varying the toroidal spectral density.

In view of the above, it is therefore possible to realistically study
the importance of the self-interaction mechanism via a flux-tube
simulation for a given size tokamak, characterized by a finite
$\rho^\star$ value, by considering the appropriate toroidal spectral
density, {\it i.e.} by setting $k_{y,\min}\rho_i =
\rho_i/C_y=(q_0a/r_0)\rho^\star$ corresponding to the toroidal mode
number $n=1$ (one naturally recovers here that the number of modes
contributing to turbulence scales as $N_\varphi\simeq
1/k_{y,\min}\rho_i\sim 1/\rho^\star$). This is clearly equivalent to
setting $L_y= 2\upi C_y$, {\it i.e.}  $\Delta\varphi=2\upi$, which along with parallel box length $L_z=2\upi$ implies having the flux-tube cover the {\it full} magnetic
surface once. Therefore, considering a finite
$k_{y,\min}\rho_i$ value in a flux-tube simulation effectively
corresponds to accounting for a finite $\rho^\star$ effect. All other
finite $\rho^\star$ effects such as profile shearing
\citep{Waltz1998,Waltz2002}, or finite radial extent of the unstable
region \citep{Ben2010}, are however obviously absent in a flux-tube.

Given that $k_{y\min}\rho_i \sim\rho^\star$, the true flux-tube model
would require considering the limit of $k_{y\min}\rho_i \to 0$. In
numerical simulations however, for obvious practical reasons,
$k_{y\min}\rho_i\sim\rho_i/L_y$ is always finite, so that if
turbulence actually converges for $k_{y\min}\rho_i \to 0$, this limit
can at best be approximately approached in the limit of large box
length $L_y/\rho_i$, which may become numerically prohibitive. Thus,
carrying out the above mentioned $k_{y,\min}\rho_i$ scan also enables
us to investigate whether/how the flux-tube simulation results
converge as $k_{y,\min}\rho_i\rightarrow 0$, a problem that was already addressed in our sister paper \citep{Justin2020}. It is remarkable
that, to our knowledge, the convergence of flux-tube gyrokinetic
turbulence simulations with respect to $k_{y\min}\rho_i$ seems not to
have been given more attention in the literature, at least not considering fully kinetic electron dynamics. Our simulations of ITG-driven
turbulence indeed illustrate that for practical values of
$k_{y,\min}\rho_i$, typically chosen in the range $10^{-2}-10^{-1}$,
convergence of turbulent fluxes is in many cases not yet
reached. Particularly in the case of strong background temperature
gradients, \emph{i.e.} away from marginal stability, the gyro-Bohm
normalised ion heat flux $Q_i$ keeps on increasing with a nearly
algebraic scaling $Q_i \sim (k_{y,\min}\rho_i)^\alpha$, $\alpha > 0$,
when decreasing $k_{y,\min}\rho_i$, thus showing no apparent sign of
convergence within this range of values. In the particular strong
gradient case considered, corresponding to parameters near to the
Cyclone Base Case (CBC) \citep{Dimits2000}, one has $\alpha \simeq
0.45$ (see Fig. \ref{Qivskymin} in this paper), so that $Q_i$
increases by nearly a factor of $3$ over the range $k_{y,\min}\rho_i =
10^{-2}-10^{-1}$. In agreement with related work by one of our
co-authors \citep{Justin2020}, for sufficiently small values of
$k_{y,\min}\rho_i$, the algebraic scaling is ultimately broken and
convergence of the fluxes is finally approached within $\sim 5\%$ for
$k_{y,\min}\rho_i \simeq 5\cdot 10^{-3}$ (see
Fig. \ref{Qivskymin_4_dxb4}). But this corresponds to a value of
$k_{y,\min}\rho_i$ one order of magnitude smaller then usually
considered. For gradients closer to marginal stability, this
dependence of $Q_i$ on $k_{y,\min}\rho_i$ appears to be weakened and
convergence approached already for somewhat larger, numerically more
affordable values. This can be seen as good news, as gradients near
marginal stability may be considered as the physically most relevant
cases. It nonetheless appears essential that one keeps in mind this
potentially significant dependence of the fluxes on $k_{y,\min}\rho_i$
in the range of typically considered values, which seems to have very
often been overlooked or at least not been given sufficient attention
in past studies.

In conjunction with the observed increase in fluxes with system size over the considered range of $k_{y,\min}\rho_i$=$10^{-2}-10^{-1}$, a decrease in the shearing rate associated to fluctuating zonal flows is also observed [see Fig. \ref{omegaeffvskymin}(c)]. Via a simple ``back-of-the-envelope'' estimate, this decrease in the shearing rate associated to fluctuating zonal flows is shown to result from the fact that the self-interaction drive of zonal flows from the various microturbulence modes are decorrelated with each other.

The remainder of this paper is organized as follows. 

Important aspects of the flux-tube version of the grid-based
gyrokinetic code GENE \citep{GENE1,GENE3,GENE2}, which was used for
performing the simulations in this study, are presented in \S
\ref{secGENE}. Details of its field-aligned coordinate system are
recalled, in particular how the parallel boundary conditions ensure
that linear eigenmodes correctly ``feel'' associated MRSs.

A brief summary of a set of previous studies \citep{Dominski2015,
  Dominski2017}, addressing the role of non-adiabatic passing electron
dynamics and which provided the starting point for the main work
presented in this paper, is given in \S \ref{Secrecap}. The same
ITG-driven turbulence case as in \citep{Dominski2015} will in fact
also be considered as a reference case here. Its parameters are
recalled as well as the linear frequency and growth rate spectra. Also
pointed out are the fine radial structures that develop on the linear
eigenmodes as a result of the non-adiabatic passing electron
dynamics. Furthermore summarized is how these fine structures on the
eigenmodes act as channels for electron heat and particle transport,
and, through non-linear coupling, are observed to lead to corrugations
at low order MRSs of the time-averaged density and temperature
profiles as well as $E\times B$ zonal flow shear layers.
%% In Sec.\ref{Secrecap}, a short summary of the effects of non-adiabatic
%% passing electron response already pointed out in
%% references~\citep{Dominski2015, Dominski2017}, is given. We use linear
%% simulations to demonstrate the formation of fine-structures at MRSs
%% corresponding to ITG eigenmodes. In non-linear simulations we
%% demonstrate MRSs acting as channels for the transport of non-adiabatic
%% passing electrons and the formation of corrugations on density and
%% temperature profiles, as well as $E\times B$ zonal flow shear layers
%% at low order MRSs.

The main results for the present study, consisting of non-linear
flux-tube simulations of ITG-driven turbulence are presented in
\S \ref{Simsetup}. Sets of simulations have been obtained by
scanning $k_{y,\min}\rho_i$ over the typical range of values
considered in practice, {\it i.e.}  $k_{y,\min}\rho_i =
10^{-2}-10^{-1}$. These scans were repeated considering ion
temperature gradients both near and far from marginal
stability. Furthermore, to clearly illustrate how the fine radial
structures resulting from the non-adiabatic passing electron dynamics
lead to particularly strong contributions to zonal flow drive from the
self-interaction mechanism, simulations with two different electron
models were carried out: either considering fully kinetic electron
dynamics or enforcing their fully adiabatic response (fine radial
structures being absent with the latter model). First analysis of
results are carried out in this same section, showing that especially
in the case of fully kinetic electrons and far from marginal
stability, heat fluxes $Q_i$ have not yet converged over the
considered range of typical $k_{y,\min}\rho_i$ values. Given its key
role in saturating ITG-driven turbulence, the level of zonal flow
shearing rate, $\omega_{E\times B}$, is carefully diagnosed. It is
shown how in all cases this shearing rate decreases with decreasing
$k_{y,\min}\rho_i$ (see figure \ref{omegaeffvskymin}). Also analyzed is
the radial correlation length of the turbulence. An important
observation is that for sufficiently small $k_{y,\min}\rho_i$, the
radial extent of turbulent eddies presents a clearly shorter scale
length then the distance between lowest order MRSs
(figure \ref{Corrvskymin}). From this one concludes that, for these
lower values of $k_{y,\min}\rho_i$, the turbulence is mostly sheared
by the zonal flows between lowest order MRSs, where $\omega_{E\times
  B}$ is essentially composed of its fluctuating component, rather
then sheared by the stationary component of $\omega_{E\times B}$
located at the lowest order MRSs.

%% We then proceed to carry out the scan in $k_{y,\min}\rho_i$ and show
%% the preliminary results on scaling of zonal flow shearing rate and
%% heat flux levels in Sec.\ref{Simsetup}.

Section \ref{AnalyzingZFdrive} is dedicated to understanding why
fluctuating zonal shear flow level decreases with 
$k_{y,\min}\rho_i$. To this end, the different zonal flow driving
mechanisms are analyzed in detail. After showing in \S \ref{ZFproxy}
that Reynolds stress can be considered as a proxy for measuring the
drive of zonal flows, we review the two basic zonal flow driving
mechanisms: modulational instability in \S \ref{Refslab} and
self-interaction in \S \ref{Reftoroidal}. To illustrate these two
basic mechanisms, reduced non-linear simulations are presented in
\S \ref{Refreduced}, where the drive of zonal modes via the decay of
an initially single finite amplitude ITG eigenmode is studied. These
simulations are carried out with both adiabatic and kinetic electrons
to demonstrate the dominant role of modulational instability in the
former case and the significant role of self-interaction in the
latter.

Evidence of the self-interaction and the modulational instability
mechanisms in fully developed turbulence is then addressed in
\S \ref{Sec. SI in turbulent sims} and \S \ref{Sec. MI in
  turbulent sims}, respectively. A comparison between simulations with
adiabatic and kinetic electrons is done here as well. It is shown that
the self-interaction mechanism is persistent in the turbulence
simulations accounting for fully kinetic electrons, while it is weak
in simulations with adiabatic electrons, consistent with the reduced
nonlinear simulations. Using statistical methods such as bicoherence
and Reynolds stress correlation analysis diagnostics, we show that the
self-interaction contributions to Reynolds stress from the various
microturbulence modes are decorrelated in time, and essentially act as
random kicks on the zonal modes, while the contributions from
modulational instability, which is a coherent process, are correlated
with each other. In the case of adiabatic electron simulations, strong
correlation between Reynolds stress contributions from different modes
as well as large bicoherence levels are measured, reflecting that
self-interaction is indeed weak and zonal flow drive is dominated by
the modulational instability. In the case of kinetic electrons
however, self-interaction may disrupt the zonal flow drive from
modulational instability and, consequently, correlation between
Reynolds stress contributions as well as bicoherence levels are found
to be relatively weak (see figures \ref{Energy_bicoh} and
\ref{normcorr}).

Based on the study carried out in \S \ref{AnalyzingZFdrive},
providing evidence of the decorrelated nature of the self-interaction
contributions to zonal flow drive from the various microturbulence
modes, we carry out a ``back-of-the-envelope'' derivation, using
simple statistical arguments, to predict the scaling observed in
simulations with kinetic electrons of the zonal flow shearing rate
$\omega_{E\times B}$ with $k_{y,\min}\rho_i$. This derivation is
presented in \S \ref{Statdep}.

%% {\bf move this paragraph somewhere else?} In all cases however, both
%% far and near marginal stability, the standard deviation over time of
%% shearing rate associated to zonal flows at any given radial position
%% is found to decrease with decreasing $k_{y,\min}\rho_i$. Based on
%% simple statistical arguments, we show that this could be a consequence
%% of the decorrelated nature of the self-interaction drive from the
%% various microturbulence modes. %\\\\

%% Using simple statistical arguments, in Sec. \ref{Statdep} we discuss
%% how the random nature of drive of zonal flows from the
%% self-interaction mechanism may explain the decrease in the shearing
%% rate associated to fluctuating zonal flows, with decreasing $k_{y,
%%   \min}\rho_i$ in kinetic electron simulations.

Final discussion and conclusions are provided in \S4 \ref{Conclusions}.
%
%----------------------------------------------------------------------
%
\section{The GENE flux-tube model}\label{secGENE}

GENE is an Eulerian electromagnetic gyrokinetic code. The flux-tube version of
the code is used in this study. In view of the issues addressed in this paper,
we recall some important features of the flux-tube model (scaling assumptions,
field-aligned coordinate system, boundary conditions).
  
GENE uses a non-orthogonal, field-aligned coordinate system $(x, y, z)$
defined in terms of the magnetic coordinates $(\psi, \chi, \varphi)$ as
follows \citep{Beer1995} :
\begin{flalign}
\label{def. x coordinate}
  x & = x(\psi) \ \ \text{: radial coordinate,} \\ 
  \label{def. y coordinate}
  y & = C_y[q_s(\psi)\chi-\varphi] \ \ \text{: binormal coordinate,} \\ 
  \label{def. z coordinate}
  z & = \chi \ \ \text{: parallel coordinate.}
\end{flalign}
$\psi, \chi$ and $\varphi$ represent the poloidal magnetic flux, straight
field line poloidal angle and the toroidal angle respectively. The function
$x(\psi)$ is a function of $\psi$ with units of length. $C_y=r_0/q_0$ is a
constant, where $q_0$ is the safety factor at $r_0$ denoting the radial
position of the flux-tube. The inverse aspect ratio of the flux-tube is
defined as $\epsilon=r_0/R$, where $R$ is the major radius of the tokamak.
  
The flux-tube model assumes a scale separation between the radial correlation
length of turbulent eddies ($\sim\rho_i$) and the radial length scale of variation of
equilibrium ($\sim a$), thus corresponding to the limit $\rho^*=\rho_i/a \ll
1$. Consistent with this scale separation, the background density and
temperature gradients, as well as the magnetic equilibrium quantities, are
considered constant across the radial extension $L_x$ of the flux-tube, and
are evaluated at $r_0$. An exception is the safety factor appearing in the
parallel boundary condition, discussed in detail later in this section. The
background density and temperature of a species $j$ are, respectively,
$N_{j,0}=N_{j,0}(r_0)$ and $T_{j,0}=T_{j,0}(r_0)$ and their inverse radial
gradients lengths are $1/L_{Nj}=-d~\text{log}~N_{j,0}/dr|_{r=r_0}$ and
$1/L_{Tj}=-d~\text{log}~T_{j,0}/dr|_{r=r_0}$. The magnetic field amplitude
$B_0=B_0(z)$, Jacobian $\mathcal{J}^{xyz}=\mathcal{J}^{xyz}(z)$ and the metric
coefficients $g^{\mu\nu}(z)=\nabla\mu\cdot\nabla\nu$ where $\mu$ and $\nu$ are
the flux-tube coordinates $(x,y,z)$, depend only on the parallel coordinate.
  
The flux-tube coordinates $(x,y,z)$ satisfy the double periodic boundary
condition in a tokamak as follows.  The $\Delta\varphi$-periodicity of any
physical quantity $\mathcal{A}$, in particular fluctuations, along the
toroidal direction $\varphi$ reads:
\[
\mathcal{A}(\psi,\chi,\varphi+\Delta\varphi)=\mathcal{A}(\psi,\chi,\varphi),
\]
and translates in $(x, y, z)$ coordinates to an $L_y$-periodicity along $y$:
\begin{equation}
  \label{periodicity in y}
  \mathcal{A}(x,y+L_y,z)=\mathcal{A}(x,y,z).
\end{equation}
where $L_y=C_y\Delta\varphi$. If the flux-tube covers the full flux-surface,
$\Delta\varphi=2\upi$, else $\Delta\varphi$ is only a fraction of $2\upi$.  The
$2\upi$-periodicity in the poloidal direction $\chi$ reads:
\[
\mathcal{A}(\psi,\chi+2\upi,\varphi)=\mathcal{A}(\psi,\chi,\varphi),
\]
and translates in $(x, y, z)$ coordinates to a pseudo-periodic condition along
$z$:
\begin{equation}
  \label{pseudo-periodicity in z}
  \mathcal{A}(x,y+C_yq_s(x)2\upi, z+2\upi)=\mathcal{A}(x,y,z).
\end{equation}
This boundary condition is referred to as the parallel boundary condition
  \citep{Scott1998}. Note that in the flux-tube model in general, one can consider a periodicity in $z$ that is larger than $2\upi$ as well \citep{Beer1995, Faber2018, Justin2020}. However, typically the $2\upi$-periodicity is considered, as it is in this work.
  
Furthermore, the radial scale separation $\rho^*\ll 1$ justifies periodic
boundary conditions along x:
\begin{equation}
  \label{periodicity in x}
  \mathcal{A}(x+L_x, y, z)=\mathcal{A}(x,y,z).
\end{equation}
Given the periodicity along $x$ and $y$ expressed by equations (\ref{periodicity in
  x}) and (\ref{periodicity in y}), a Fourier series decomposition is a
practical representation of fluctuation fields as it naturally verifies these
boundary conditions. Such a Fourier representation reads:
\begin{equation}
  \label{Fourier series}
  \mathcal{A}(x, y, z) 
  = 
  \sum_{k_x, k_y}
  \hat{\mathcal{A}}_{k_x, k_y}(z)\exp[i(k_xx+k_yy)],
\end{equation}
with $k_x$ and $k_y$ corresponding in general to all harmonics of the minimum
wave numbers $k_{x, \min} = 2\upi/L_x$ and $k_{y, \min} = 2\upi/L_y$
respectively.
  
The underlying axisymmetry of a tokamak corresponds to an invariance of the
unperturbed system with respect to $\varphi$ in $(\psi, \chi, \varphi)$
coordinates. This translates to an invariance with respect to $y$ in $(x,y,z)$
coordinates. Consequently, any fluctuating field related to a {\it linear}
eigenmode of the toroidal system is thus composed of a single $k_y$ Fourier
mode. Note that a given $k_y$ wave number is equivalent
to a toroidal wave number $n$ according to the relation
\begin{equation}
  \label{relation between ky and n}
  n=-k_yC_y.
\end{equation}
  
It remains to express the pseudo-periodic boundary condition
(\ref{pseudo-periodicity in z}) in the Fourier representation (\ref{Fourier
  series}). In doing so, one considers the $x$-dependence of the safety factor
profile $q_s(x)$. This is essential to ensure that the flux-tube model contains
the information on the radial position of MRSs related to a $k_y$ mode and
thus accounts for the particular dynamics, in particular resonances, that can
take place at such surfaces. This is of key importance to the study carried
out in this paper. Accounting for the $x$-dependence of the safety factor in
these boundary conditions is thus an exception in the flux-tube framework, as
all other background and magnetic geometry coefficients, as already mentioned,
are assumed $x$-independent. Only a {\it linearised} safety factor profile of
the form:
\begin{equation}
  \label{linearized q profile}
  q_s(x)=q_0[1+\hat{s}(x-x_0)/r_0],
\end{equation}
with $\hat{s}$ standing for the magnetic shear, is in fact compatible with the
Fourier representation along $x$. For a given $k_y$ Fourier mode, inserting
(\ref{Fourier series}) and (\ref{linearized q profile}) into
(\ref{pseudo-periodicity in z}) leads to
\begin{equation}
  \label{pseudo-periodicity in z, Fourier, V1}
  \sum_{k_x}
  \hat{\mathcal{A}}_{k_x, k_y}(z+2\upi)
  \exp[2\upi i\, k_yC_yq_s(x=0)] \exp[i (k_x + 2\upi k_y\hat{s})x]
  =
  \sum_{k_x}
  \hat{\mathcal{A}}_{k_x, k_y}(z)
  \exp(i k_x x).
\end{equation}
  
For convenience and without further loss of generality, one assumes here that
the origin of the $x$ coordinate corresponds to the lowest order MRS (LMRS), {\it i.e.}
a MRS related to $k_{y, \min}$. This condition reads:
\[
k_{y, \min}C_yq_s(x=0) 
=
n_{\min} q_s(x=0) 
= 
n_{\min} q_0(1-\hat{s}x_0/r_0) \;\in \mathbb{Z},
\]
with $n_{\min}$ the toroidal wave number associated to $k_{y, \min}$.  Note
that this relation provides an equation for the shift in $x_0$ and ensures
that the phase factor $\exp[2\upi i\, k_yC_yq_s(x=0)]=1$ {\it for all} $k_y$
modes. Based on (\ref{pseudo-periodicity in z, Fourier, V1}), the boundary
condition in $z$ then finally translates for a given $k_y$ Fourier mode to the
following coupling between $k_x$ Fourier modes:
\begin{equation}
  \label{pseudo-periodicity in z, Fourier, V2}
  \hat{\mathcal{A}}_{k_x, k_y}(z+2\upi) 
  =
  \hat{\mathcal{A}}_{k_x+2\upi k_y\hat{s},\,k_y}(z). 
\end{equation}
  
The coupling between $k_x$ Fourier modes can also be understood as follows. In
a sheared toroidal system, the wave vector associated to a Fourier mode $(k_x,
k_y)$ is given by
\begin{equation}
  \label{wave vector of mode (kx, ky)}
  \vec{k} 
  = 
  k_x\nabla x + k_y \nabla y 
  = 
  (k_x + k_y\,\hat{s}\,z)\nabla x 
  - nq_s\nabla\chi 
  + 
  n\nabla\varphi,
\end{equation}
having used relations (\ref{def. y coordinate}) and (\ref{relation between ky
  and n}). After one poloidal turn ($z\to z+2\upi$), the wave vector (\ref{wave
  vector of mode (kx, ky)}) obviously becomes the one associated to the
Fourier mode $(k_x+2\upi k_y\hat{s},\,k_y)$, thus explaining the coupling of
$k_x$ modes.
    
One should emphasize that this coupling resulting from the parallel boundary
condition applies to any fluctuating field and in particular already to linear eigenmodes of the system and is thus of different physical nature than the three Fourier mode ($\sim$ 3-wave) interaction discussed later on, resulting from non-linear dynamics.
  
A linear eigenmode with fixed $k_y$ thus couples to a set of $k_x = k_{x0} +
p\,2\upi k_y\hat{s}$, $p\in\mathbb{Z}$, modes and is of the form:
\begin{equation}
  \label{linear eigenmode, Fourier rep.}
  \mathcal{A}(x, y, z) 
  = 
  \exp(ik_yy)
  \sum_{p=-\infty}^{+\infty}
  \hat{\mathcal{A}}_{k_{x0} + p2\upi k_y\hat{s},\,k_y}(z)
  \exp[i(k_{x0} + p\,2\upi k_y\hat{s})x].
\end{equation}
One can show that this form is equivalent to the so-called ballooning
representation \citep{Connor1978,Hazeltine1990}:
\begin{equation}
  \label{ballooning rep.}
  \mathcal{A}(\psi, \chi, \varphi) 
  = 
  \sum_{p=-\infty}^{+\infty}
  \hat{\mathcal{A}}_b(\chi+p\,2\upi)
  \exp[in(\varphi -q_s(\psi)(\chi+p\,2\upi-\chi_0))],
\end{equation}
noting in this relation the ballooning envelope $\hat{\mathcal{A}}_b(\chi)$,
defined over the extended ballooning space $\chi\in]-\infty, +\infty[$, as
    well as the field-aligned phase factor including the ballooning angle
    $\chi_0$. The relation between the two representations (\ref{linear
      eigenmode, Fourier rep.}) and (\ref{ballooning rep.}) is given by
\begin{flalign}
  \label{ballooning defs.}
  \hat{\mathcal{A}}_b(\chi+p\,2\upi) 
  & = 
  \hat{\mathcal{A}}_{k_{x0} + p2\upi k_y\hat{s},\,k_y}(\chi)
  \end{flalign}
\begin{flalign}
\label{kx0deff}
  \chi_0 
  & = 
  -k_{x0}/(k_y\hat{s}),
\end{flalign}
  
In a flux-tube of radial extension $L_x$, all coupled Fourier modes $k_x +
p\,2\upi k_y \hat{s}$ relative to this direction must be harmonics of $k_{x,
  \min}$. This must hold for all $k_y$ and in particular for the lowest
harmonic $k_{y, min}$:
\[
2\upi k_{y, \min} \hat{s} 
= 
M\,k_{x, min} 
= M\,\frac{2\upi}{L_x},
\]
with $M \in \mathbb{N}^\star$ a strictly positive integer. This relation can
be rewritten:
\begin{equation}
  \label{relation between Lx and Ly}
  L_x 
  = 
  \frac{M}{k_{y, \min}\hat{s}}
  = 
  M\,\Delta x_{\rm LMRS} 
  = 
  \frac{M}{2\upi\hat{s}}\,L_y,
\end{equation}
thus imposing a constraint between the extensions $L_x$ and $L_y$ of the flux
tube along the directions $x$ and $y$ respectively. In practice, the integer
$M$ must be chosen such that $L_x$ is larger then the radial correlation
length of turbulent eddies.
  
Relation (\ref{relation between Lx and Ly}) also implies that $L_x$ must be an
integer multiple of $\Delta x_{\rm LMRS} = 1/(k_{y, \min}\hat{s})$, identified
as the distance between lowest order MRSs. Indeed, considering the linearised
safety factor profile (\ref{linearized q profile}), the distance $\Delta
x_{\rm MRS}$ between MRSs corresponding to a given $k_y\neq 0$ mode is
constant and given by
\[
1 = \Delta [n q_s(x)] = k_y\hat{s}\,\Delta x_{\rm MRS}
\quad\Longrightarrow\quad
\Delta x_{\rm MRS}(k_y) = 1/(k_y\hat{s}),
\]
having invoked (\ref{relation between ky and n}). One thus in particular has
$\Delta x_{\rm LMRS} = \Delta x_{\rm MRS}(k_{y,\min}) = 1/(k_{y,
  \min}\hat{s})$.  For a given $k_y\neq 0$ mode, the radial positions of
corresponding MRSs are thus
\[
x_{\rm MRS} 
= 
m\,\Delta x_{\rm MRS} 
= 
m\,\frac{k_{y, \min}}{k_y}\,\Delta x_{\rm LMRS}, 
\quad m\in\mathbb{Z}.
\]
Note that the positions of lowest order MRSs, $x_{\rm LMRS} = m\,\Delta x_{\rm
  LMRS}$, are MRSs to {\it all} $k_y\neq 0$ modes. More generally, there tends
to be an alignment of the radial positions of MRSs corresponding to the
different $k_y$ around the lowest order MRSs (second order MRSs are common to
every second $k_y$, third order MRSs to very third $k_y$, etc.)  and a
misalignment around the higher order MRSs, as shown in
figure~\ref{recap}(a). This level of (mis)alignment of MRSs can be measured by
their radial density, as shown in figure 12 in \citep{Dominski2017}.

Other numerical parameters are as follows. In the $k_x$ and $k_y$ Fourier spaces relative to the $x$ and $y$ directions, one considers $N_{k_x}$ and
$N_{k_y}$ Fourier modes respectively. In real space, the simulation volume
$L_x\times L_y\times L_z$ is discretised by $N_x\times N_y\times N_z$
equidistant grid points such that $N_x=N_{k_x}$ and $N_y=2N_{k_y}-1$. Note that by invoking the reality condition, only modes $k_y\geq 0$ are evolved in GENE. The parallel direction is treated in real space with $z \in [-L_z/2, +L_z/2[$. For our study, which in particular addresses a system size effect, it is essential to choose $L_z=2\upi$, {\it i.e.} considering only one poloidal turn, for consistency with global geometries \citep{Scott1998}.
The gyrocenter velocity space coordinates are $(v_{\parallel},\mu)$, where
$\mu=mv_{\perp}^2/2B_0$ is the magnetic moment, and $v_\parallel$ and
$v_{\perp}$ are respectively the velocity components parallel and
perpendicular to the magnetic field. For the parallel velocity direction
one considers $v_\parallel \in [-v_{\parallel, \max}, +v_{\parallel,
\max}]$ with a discretisation involving $N_{v_\parallel}$ equidistant grid
points, while for the $\mu$ direction one considers $\mu \in [0,
\mu_{\max}]$ with a discretisation involving $N_\mu$ grid points following the
Gauss-Laguerre integration scheme.

%
%----------------------------------------------------------------------
%
 \section{Non-adiabatic passing electron dynamics leading to stationary zonal structures}\label{Secrecap}
 
 This section presents a summary of relevant results from the articles
  \citep{Dominski2015,Dominski2017}, which addressed certain effects of non-adiabatic
  passing electron dynamics on turbulent transport.
  
\begin{figure}
\centering
  \includegraphics[width=1\linewidth]{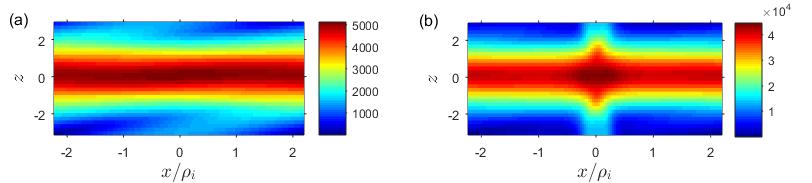}
  \caption{Linear eigenmode with $k_{x0}=0$ and $k_y\rho_i=0.28$, in the case
    of (a) adiabatic, (b) kinetic electron response. Shown is the $(x,
    z)$-dependence of the electrostatic potential $\Phi$ in absolute value,
    weighted by the Jacobian, $J |\Phi|$.}
  \label{phionxandz}
\end{figure}

\begin{figure}
\centering
  \includegraphics[width=0.51\linewidth]{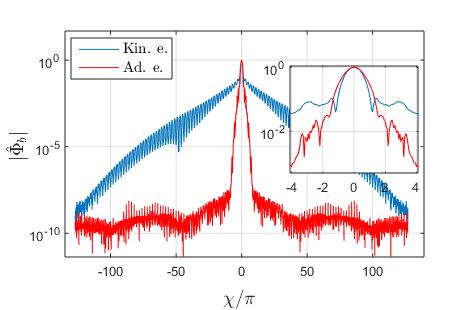}
  \caption{Ballooning representation $|\hat{\Phi}_b(\chi)|$ of the
    electrostatic potential $\Phi$ for the same linear eigenmode as in
    figure~\ref{phionxandz}, showing both the case of kinetic (blue) and
    adiabatic (red) electrons. Inset figure shows the zoom near
    $\chi=0$.}
  \label{Ballooning}
\end{figure}

\begin{figure}
\centering
  \includegraphics[width=0.44\linewidth]{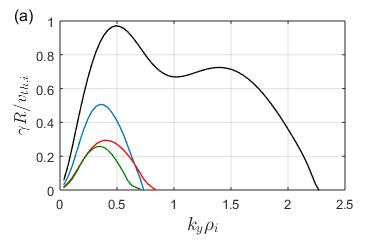}
  \includegraphics[width=0.44\linewidth]{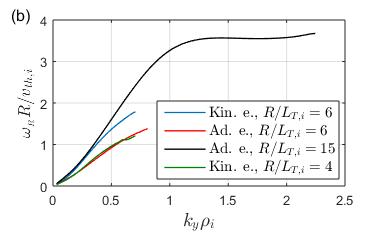}  
  \caption{(a) Linear growth rate $\gamma$ and (b) real frequency $\omega_R$
    in units of $v_{{\rm th},i}/R$ as a function of $k_y\rho_i$. Adiabatic electron
    simulations with $R/L_{T,i}=6$ and $15$ are plotted in red and black
    respectively. Kinetic electron simulations with $R/L_{T,i}=4$ and $6$ are
    plotted in green and blue respectively.}
  \label{gammavsky_kinadcomp}
\end{figure}

  Non-adiabatic passing electron response leads to generation of fine-structures at MRSs which can
  be first studied in linear simulations. Figure~\ref{phionxandz}
  shows the envelope in the $(x, z)$ plane of the electrostatic potential
  $\Phi$ of the unstable linear ITG eigenmode with $k_y\rho_i=0.28$,
  considering either adiabatic or kinetic electron response, and for the set
  of physical parameters given in Table~\ref{Parameterset}. This is the same
  ITG case as considered in reference \citep{Dominski2015} and same numerical grid
  resolutions have been considered here. While in both cases the modes are
  ballooned at $z=0$, one observes a fine radial structure at the
  corresponding MRS (positioned at the center $x=0$ of the radial domain) only
  in the latter case. This is the result of the non-adiabatic passing electron
  response taking place at MRSs where the parallel wavenumber
  $k_\parallel\rightarrow 0$ \citep{Chowdhury2008,Dominski2015}. In the
  vicinity of MRSs, the condition for adiabatic electron response is violated
  as the phase velocity of the eigenmode parallel to the magnetic field
  becomes greater than the electron thermal velocity:
  $|\omega/k_\parallel|>v_{{\rm th},e}$. These fine structures along the $x$ direction translate to a broad range
  of significant $k_x$ Fourier modes, \emph{i.e.} to a broad ballooning
  structure according to equation~(\ref{ballooning defs.}), which is referred to
  as the ``giant electron tails'' in \citep{Hallatscheck2005}; see figure~\ref{Ballooning}. No such broad
  tails in the ballooning structure are observed with adiabatic electrons.
  Figure~\ref{gammavsky_kinadcomp} plots the
  $k_y$ spectrum of linear growth rates $\gamma$ and real frequencies
  $\omega_R$ of most unstable eigenmodes for the cases considered here. Note that $\omega_R>0$ corresponds
  in GENE convention to a mode propagating in the ion-diamagnetic direction,
  as expected for ITG instabilities.
  
\begin{figure}
\centering
  \includegraphics[width=1\linewidth]{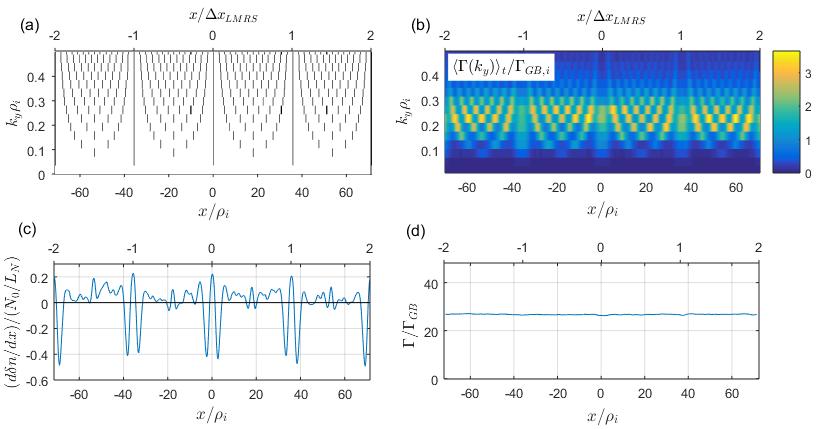}
  \caption{(a) Radial position of MRSs for each $k_y$ mode, (b) radial
    dependence of contribution to time-averaged particle flux $\Gamma$ in
    gyro-Bohm units $\Gamma_{GB}=N_0v_{{\rm th},i}(\rho_i/R)^2$ from each $k_y$
    mode, (c) radial gradient of flux-surface and time-averaged density
    fluctuation $\delta n$ normalised with respect to $N_0/L_N$
    (positive/negative values correspond resp. to flattening/steepening of
    profiles), and (d) radial profile of time-averaged total particle flux (summed over all $k_y$). 
    All subplots correspond to kinetic electron simulation for the 
    parameter set given in table~\ref{Parameterset}, with $k_{y, \min}\rho_i=0.035$ and $R/L_{T,i}=6$.
    Ticks $x/\Delta x_{LMRS}=\{-2,-1,0,1,2\}$ 
    on the top axes denote the lowest order MRSs (LMRSs). }
  \label{recap}
\end{figure}

  The radial structures related to the non-adiabatic passing electron
  dynamics have been shown to persist in the non-linear turbulent regime,
  as discussed in references \citep{Waltz2006,Dominski2015,Dominski2017}. Studies by \citet{Dominski2015,Dominski2017}, based on both local (flux-tube) and
  global gyrokinetic simulations, have furthermore shown that, for each
  fluctuation mode with mode number $k_y$, the corresponding MRSs act as
  radially localized transport channels. This is
  illustrated in figures~\ref{recap}(a,b). Radial regions with high (resp. low)
  density of MRSs thus tend to lead to high (resp. low) particle and heat
  diffusivities. Consequently, to ensure radially constant time-averaged total
  particle and heat fluxes, density and temperature gradients (driving the
  turbulence) become corrugated, steepening in regions with low density of
  MRSs and flattening in regions with high density of MRSs. See
  figures~\ref{recap}(c,d) where the time-averaged density gradient and particle flux are shown as a function $x$.
  
\begin{figure}
\centering
\includegraphics[width=0.72\columnwidth] {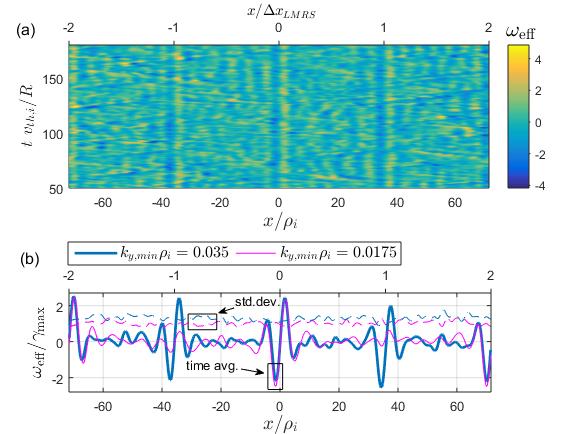}
  \caption{(a) Effective shearing rate $\omega_{\rm eff}(x, t)$ plotted as a
    function of the radial position $x$ and time $t$, for the kinetic electron simulation with parameters as given in table~\ref{Parameterset}, $k_{y, \min}\rho_i=0.035$ and $R/L_{T,i}=6$. (b) Solid blue line  indicates the associated
    time-averaged component $\langle\omega_{\rm eff}\rangle_t$, normalised by the corresponding maximum linear growth rate $\gamma_{\max}$, at each radial position
    $x$. Dashed blue line indicates the standard
    deviation ${\rm SD}_t(\omega_{\rm eff})(x) = \left[\, \langle\, \left(
      \omega_{\rm eff} - \langle\omega_{\rm eff}\rangle_t \right)^2
      \,\rangle_{t} \,\right]^{1/2}$ in time, normalised by $\gamma_{\max}$. For comparison, results for $k_{y, \min}\rho_i=0.0175$ have been added
    in magenta.}
  \label{omegaecrossbionsvsx_sepspcompare}
\end{figure}

  Stationary $E\times B$ shear flow layers associated to the
  time-averaged radial electric field are also observed (see
  figure~\ref{omegaecrossbionsvsx_sepspcompare} where the
  corresponding effective shearing rate as defined in
  (\ref{effsheardef}) is plotted). This electric field component
  ensures the radial force balance with the pressure gradients related
  to the corrugation of density and temperature profiles
  \citep{Waltz2006}. In sections \ref{Reftoroidal} - \ref{Sec. SI in
    turbulent sims}, it is shown that these stationary shear
  structures are actually driven by a contribution to the Reynolds
  Stress coming from the so-called self-interaction mechanism
  \citep{Weikl2018}.
  
  %{\color{blue}: The fine scale structures that develop on linear eigenmodes at corresponding MRSs as a result of the non-adiabatic passing electron dynamics can non-linearly interact with themselves to provide a specific contribution to the RS. At lowest order MRSs these contributions from the different $k_y$ modes are spatially aligned, thus leading to the observed time-averaged shear layers. Away from lowest order MRSs, these contributions are however misaligned and on average over time cancel out, thus leading to the absence of stationary zonal flow structure in these regions (as will be illustrated in figure~\ref{d2RS}).{\bf Remove this section?}}
  
%
%----------------------------------------------------------------------
%
\section{$k_{y, \min}\rho_i$ scan in ITG-driven turbulence. Role of stationary
  and fluctuating components of zonal shear flows}\label{Simsetup}

Non-linear simulations have been carried out considering conditions of
ITG-driven turbulence. Reference physical and numerical parameters for
these simulations are summarized in table~\ref{Parameterset}. The
physical parameters are close to the Cyclone Base Case (CBC)
\citep{Dimits2000}, with background gradients slightly modified to
eliminate unstable TEM and ETG modes. Compared to typical flux-tube
runs, a high radial resolution is chosen (with radial grid-points
$N_x=512$) in order to ensure that the dynamics at MRSs (separated by
a radial distance $\Delta x_{MRS}=1/\hat{s}k_y$ for a given $k_y$) is
very well resolved, as discussed in detail in
Ref. \citep{Dominski2015}. In our work, this high $x$-resolution is
all the more critical since the fine-structures forming at the MRSs
play an important role in the self-interaction mechanism, which is
crucial to our study.
  
A scan in $k_{y, \min}\rho_i$ is performed while keeping both $k_{y,
  \max}\rho_i$ and $L_x/\rho_i$ fixed. Successively halving $k_{y,
  \min}\rho_i$, the values $k_{y, \min}\rho_i = 0.14$, $0.07$,
$0.035$, and $0.0175$ have been considered. Note that these values for
$k_{y, \min}\rho_i$ span the typical range $10^{-2}-10^{-1}$
considered in practice when carrying out ion-scale flux-tube
simulations. To keep $k_{y, \max}\rho_i$ fixed, the total number $N_y$
of $k_y$ modes must thus be doubled between consecutive runs, whereas
to keep $L_x/\rho_i$ fixed, the number $M$ of lowest order MRSs
contained in the simulation box is halved. The parameter $M$ thus
takes on the respective values $M= 16$, $8$, $4$ and $2$.  The case
$k_{y, \min}\rho_i=0.07$ is in fact equivalent to the ITG case already
studied by \citet{Dominski2015}.
  
  \begin{table}
\begin{footnotesize}
\begin{tabular}{ p{2.3cm} p{2.3cm} p{2.1cm} p{2.9cm} p{2.9cm}  }
  \multicolumn{5}{l}{Geometry: Ad-hoc concentric circular
    geometry\citep{Lapillonne2009}} \vspace{0.1cm}
  \\ 
  $\epsilon=0.18$ & $q_0=1.4$ & $\hat{s}=0.8$  & $\beta=0.001$ & \vspace{0.2cm}
  \\ 
  $m_i/m_e=400$ & $T_{e}/T_{i}=1.0$ & $R/L_{N}=2.0$ & $R/L_{T_{i}}=6.0$ & 
  $R/L_{T_{e}}=2.0$ \vspace{0.15cm}
  \\ 
  $L_x=142.9\,\rho_i$ & $L_y^*=179.5\,\rho_i$ & $L_z = 2\upi$ & $v_{\parallel, \max} =
  3\sqrt{2}\,v_{{\rm th}, i}$ & $\mu_{\max}=9\,T_i/B_{0, {\rm axis}}$ \vspace{0.2cm}
  \\ 
  $M^*=4$ & \multicolumn{4}{l}{$N_{k_x} \times N_{k_y}^* \times N_z \times
    N_{v_{\parallel}} \times N_\mu = 512 \times 128 \times 16 \times 64 \times
    9$} 
  \\ 
\end{tabular}
\end{footnotesize}
\caption{Parameter set for non-linear simulations. The parameter $k_{y, \min} =
  2\upi/L_y$ is scanned and takes the values $k_{y, \min}\rho_i\in\{0.14,
  0.07, 0.035, 0.0175\}$, while $k_{y,\max}=k_{y, \min}N_y/2$ and
  $L_x=M/\hat{s}k_{y, \min}$ are kept constant. The values indicated in
  the table correspond to the particular case
  $k_{y, \min}\rho_i=0.035$. Asterisks indicate variables which vary during the $k_{y, \min}\rho_i$
  scan. $v_{{\rm th}, i} = \sqrt{T_i/m_i}$ stands for the ion thermal velocity and
  $B_{0, {\rm axis}}$ for the magnetic field on axis.}
\label{Parameterset}
\end{table}        
  
  Based on the relation $k_{y,\min}\rho_i = (q_0a/r_0)\rho^*$ (assuming
  that all toroidal modes are accounted for, so that $n_{\min}=1$) and for the
  typical mid-radius value $r_0/a=0.5$ and here considered $q_0=1.4$, one
  obtains $\rho^* = 5.0\cdot 10^{-2}$, $2.5\cdot 10^{-2}$, $1.25\cdot
  10^{-2}$, $6.25\cdot 10^{-3}$. Note for reference that typical values for
  this parameter are $\rho^\star\simeq 1\cdot 10^{-2}$ in a smaller-size machine such as the TCV tokamak,
  $\rho^\star\simeq 3\cdot 10^{-3}$ in the DIII-D tokamak\citep{Waltz2005b},
  while the projected values for ITER are still an order of magnitude smaller.
  
  In order to address how the results from the $k_{y, \min}\rho_i$ scan
  depend on whether one is near or far from marginal stability, the scan was
  repeated for a second ion temperature gradient in both the adiabatic and
  kinetic electron cases. To this end, carrying out preliminary $R/L_{T_i}$
  scans for $k_{y, \min}\rho_i = 0.035$, the non-linear (Dimits-shifted) critical
  temperature gradients $R/L_{T_i, {\rm crit}}$ were first identified. For
  adiabatic electrons, $R/L_{T_i, {\rm crit}}= 5.5$ was found, so that the
  reference case temperature gradient $R/L_{T_i} = 6$ is relatively {\it near}
  marginal stability and the second $k_{y, \min}\rho_i$ scan was therefore
  performed for $R/L_{Ti}= 15$ in this case, {\it i.e.} {\it farther} from
  marginal stability. For kinetic electrons, $R/L_{T_i, {\rm crit}}= 3.5$, so
  that the reference case temperature gradient $R/L_{T_i} = 6$ is relatively
  {\it far} from marginal stability and the second $k_{y, \min}\rho_i$ scan was
  therefore performed for $R/L_{Ti}= 4$ in this case, {\it i.e.} {\it nearer}
  marginal stability.
  
  To ensure that the simulations results are sound, convergence tests with
  respect to radial box size $L_x$ and the numerical resolutions $N_z$,
  $N_{v_\|}$ and $N_\mu$ were carried out. Convergence test on $N_x$ had
  already been addressed in \cite {Dominski2015}. Based on these tests, the
  turbulent heat and particle fluxes, as well as statistical properties of
  $E\times B$ shearing rates are estimated to be within $\sim10\%$ of their
  converged value.  Benchmarking of the GENE results with the gyrokinetic code
  GS2 \citep{Dorland2000} was furthermore performed for a limited number of
  simulations. Although a reduced mass ratio is considered here, similar
  results have been obtained with the physical mass ratio of hydrogen
  $m_i/m_e=1836$.
  
  First results from the $k_{y, \min}\rho_i$ scan will now be
  discussed. Given their importance in saturating ITG turbulence, particular
  attention will be given to the statistical properties of the shearing rate
  $\omega_{E\times B}$ associated to the to zonal $E\times B$ flows. We will
  in fact consider the effective shearing rate $\omega_{\rm eff}$, similar to
  that defined in \citep{Dominski2015}. This rate is estimated as follows. One first
  defines the zonal $E\times B$ shearing rate experienced by the ions, which
  are the dominant instability drivers in the case of the here considered ITG
  turbulence:
  \begin{equation*}
  \omega_{E\times B,{\rm ion}}(x, t) 
  =
  \frac{1}{B_0}\,
  \frac{\partial^2\langle\bar{\Phi}\rangle_{y,z}}{\partial x^2},
  \end{equation*}
  where the flux-surface average $\langle\bar{\Phi}\rangle_{y,z}$ provides the
  zonal component of $\bar{\Phi}$ and involves both an average over $y$,
  $\langle\cdot\rangle_y = (1/L_y)\int_0^{L_y} \cdot\,dy$, and an average over
  $z$, $\langle\cdot\rangle_z = \int_{-\upi}^{+\upi}\cdot\,J^{xyz}
  dz/\int_{-\upi}^{+\upi} J^{xyz}dz$. $\bar{\Phi}$ is the scalar potential
  gyroaveraged over the Maxwellian ion background velocity distribution. The
  shearing rate $\omega_{E\times B,{\rm ion}}$ is then furthermore averaged
  over a small time window of width $\tau$, given that fluctuations that are
  very short-lived in time do not contribute effectively towards the zonal
  flow saturation mechanism \citep{Hahm1999}, thus providing the effective
  shearing rate:
  \begin{equation}
  \label{effsheardef}
  \omega_{\rm eff}(x, t) 
  = 
  \frac{1}{\tau}
  \int_{t-\tau/2}^{t+\tau/2} \hspace{-0.5cm}
  \omega_{E\times B,{\rm ion}}(x, t')\,dt'.
  \end{equation}
  Here, $\tau=1/\gamma_{\max}$ is considered, where $\gamma_{\max}$ is the
  growth rate of the most unstable linear mode.
  
  As an illustration, the effective shearing $\omega_{\rm eff}(x, t)$ has
  been plotted in figure~\ref{omegaecrossbionsvsx_sepspcompare}(a) as a function
  of $x$ over the full flux-tube width $L_x$ and as a function of $t$ over the
  simulation time slice $50 < t\,v_{\rm th, i}/R < 180$. Shown is the case
  with kinetic electrons, $R/L_{T_i}=6$ and $k_{y, \min}\rho_i = 0.035$. The
  radial profile for the time-averaged component $\langle\omega_{\rm
    eff}\rangle_t(x)$, is shown in figure~\ref{omegaecrossbionsvsx_sepspcompare}(b). For comparison, the same profile
  is also shown for the same physical parameters except for $k_{y, \min}\rho_i
  = 0.0175$.
  
\begin{figure}
\centering
  \includegraphics[trim=0 0 0cm 0, clip, width=0.32\linewidth]{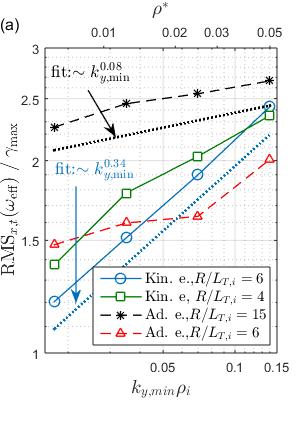}
  \includegraphics[trim=0 0 0cm 0, clip, width=0.32\linewidth]{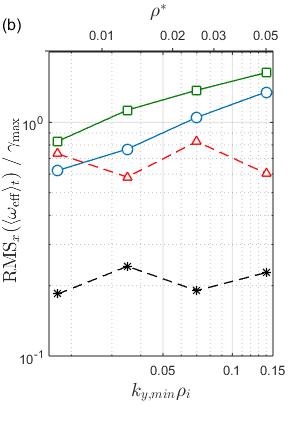}
  \includegraphics[trim=0 0 0cm 0, clip, width=0.32\linewidth]{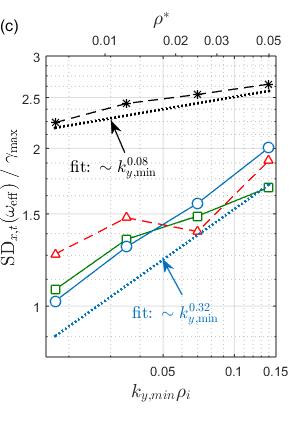}  
  \caption{Effective shearing rate $\omega_{\rm eff}$ associated to the zonal
    $E\times B$ flows, normalised to corresponding maximum linear growth rate
    $\gamma_{\max}$, as a function of $k_{y, \min}\rho_i$. Solid lines denote
    kinetic electron simulations for $R/L_{T,i}=4$ (green squares) and $6$
    (blue circles) respectively. Dashed lines denote adiabatic electron
    simulations for $R/L_{T,i}=6$ (red triangles) and $15$ (black stars)
    respectively. Other parameters remain the same as in table~\ref{Parameterset}. (a) System average of total shearing rate ${\rm
      RMS}_{x,t}(\omega_{\rm eff}) = \left(\, \langle\, \omega_{\rm eff}^2
    \,\rangle_{x,t} \,\right)^{1/2}$. (b) Contribution from the stationary
    component, ${\rm RMS}_{x}(\langle\omega_{\rm eff}\rangle_t) = \left[\,
      \langle\, \left(\, \langle\omega_{\rm eff}\rangle_t \,\right)^2
      \,\rangle_{x} \,\right]^{1/2}$. (c) Contribution from fluctuation
    component, ${\rm SD}_{x,t}(\omega_{\rm eff}) = \left[\, \langle\, \left(
      \omega_{\rm eff} - \langle\omega_{\rm eff}\rangle_t \right)^2
      \,\rangle_{x,t} \,\right]^{1/2}$. All plots in log-log scale.}
  \label{omegaeffvskymin}
\end{figure}

  The system average of the {\it total} effective shearing rate is
  plotted in figure~\ref{omegaeffvskymin}(a) as a function of $k_{y, \min}\rho_i$,
  considering a log-log scale. This average value is provided by the Root Mean
  Square (RMS) estimate:
  \begin{equation}
    \label{RMS_xt(w_eff)}
    {\rm RMS}_{x,t}(\omega_{\rm eff})
    =
    \left(\,
    \langle\,
        \omega_{\rm eff}^2
    \,\rangle_{x,t}
    \,\right)^{1/2},
  \end{equation}
  involving both a radial average, $\langle\cdot\rangle_x =
  (1/L_x)\int_0^{L_x} \cdot\,dx$, and an average over the whole
  simulation time $t_{\rm sim}$, $\langle\cdot\rangle_t = (1/t_{\rm
    sim})\int_0^{t_{\rm sim}} \cdot\,dt$.
  Figure~\ref{omegaeffvskymin}(a) shows results for the scans carried
  out for the two respective temperature gradients $R/L_{T_i}$
  considered for both adiabatic and kinetic electrons. The shearing
  rates have been normalised with respect to the value of
  $\gamma_{\max}$, which takes on different values for the four
  considered datasets (see
  figure~\ref{gammavsky_kinadcomp}). Normalised shearing values
  $\omega_{\rm eff}/\gamma_{\max} \gtrsim 1$ can be considered as
  significant for saturating ITG-driven turbulence. Note that a
  straight system average $\langle \omega_{\rm eff}(x,t)
  \rangle_{x,t}$ of the shearing rate would converge to zero over
  sufficiently long simulation time, which is why the RMS estimate
  (\ref{RMS_xt(w_eff)}) is considered.  One notes that, over the
  considered range $10^{-2} -10^{-1}$ in $k_{y,\min}\rho_i$ values,
  the total effective shearing rate ${\rm RMS}_{x,t}(\omega_{\rm
    eff})$ {\it decreases} in all cases with decreasing $k_{y,
    \min}\rho_i$, {\it i.e.} with increasing machine size. However,
  significantly stronger scaling is observed for the scans with
  kinetic compared to adiabatic electrons. As can be seen from the
  log-log plot, the shearing rate appears to roughly scale as
  $\sim(k_{y, \min}\rho_i)^\alpha$, $\alpha>0$. This scaling is
  particularly evident for the kinetic electron case far from marginal
  stability ($R/L_{T_i}=6$), for which $\alpha\simeq 0.34$. Similar
  scaling is observed nearer marginal stability ($R/L_{T_i}=4$) as
  well. The adiabatic electron scans however show a weaker scaling
  with $\alpha\simeq 0.08$ for both $R/L_{T_i}=6$ and $R/L_{T_i}=15$.

\begin{figure}
\centering
  \includegraphics[width=0.42\linewidth]{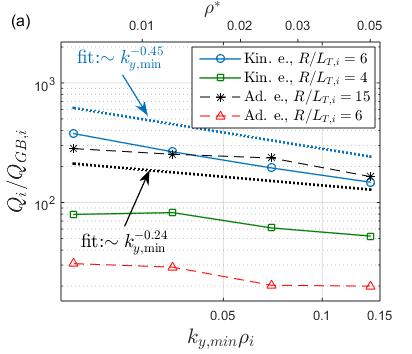}
  \hspace{0.5cm}
  \includegraphics[width=0.42\linewidth]{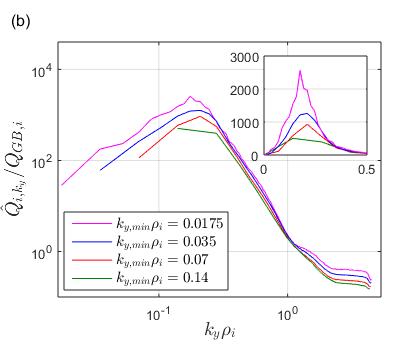}
  \caption{(a) Log-log plot of the time-averaged and gyro-Bohm
    normalised ion heat flux $Q_i$ as a function of $k_{y,
      \min}\rho_i$. Same cases as considered in
    figure~\ref{omegaeffvskymin}. (b) Log-log plot of $k_y$ spectra of
    ion heat flux $Q_i$ for the kinetic electron runs with
    $R/L_{T,i}=6$ and $k_{y, \min}\rho_i=$ 0.0175 (magenta), 0.035
    (blue), 0.07 (red) and 0.14 (green). Inset figure shows lin-lin
    plot of the zoom near the peaks.}
  \label{Qivskymin}
\end{figure}

  Figure~\ref{Qivskymin}(a) plots in log-log scale, the time and flux-tube
  -averaged radial ion heat flux $Q_i$ as a function of $k_{y,
    \min}\rho_i$. Heat fluxes have been normalised to gyro-Bohm units $Q_{GB,i} =
  n_{0, i}T_iv_{{\rm th},i}(\rho_i/R)^2$. Results for all four considered datasets are
  again presented. Over the considered range of
  $k_{y, \min}\rho_i$, fluxes appear not to be converged with respect to this
  parameter. This non-convergence is particularly striking for both adiabatic
  and kinetic electron simulations for the respective ion temperature
  gradients far from marginal stability.  As can be seen from the log-log
    plot, in these cases one observes a scaling $Q_i/Q_{GB,i} \sim
    (k_{y, \min}\rho_i)^{-\alpha}$, $\alpha>0$, with $\alpha = 0.45$ (kin.e.,$R/L_{Ti}=6$) and $0.24$ (ad.e.,$R/L_{Ti}=15$).
    
    To provide insight into which wavelengths mainly contribute to the
  increasing heat flux as $k_{y, \min}\rho_i$ decreases, the heat flux
  $k_y$-spectra is plotted in figure~\ref{Qivskymin}(b) for the set of simulations
  corresponding to the $k_{y, \min}\rho_i$ scan with kinetic electrons and
  $R/L_{T_i}=6$. $Q_{i,ky}$ is defined such that total heat flux $Q_i=\sum_{ky}Q_{i,ky}dk_y$, where $dk_y=k_{y, \min}$. One observes that all spectra present a peak at
  $k_y\rho_i\simeq 0.2$ and that the increase in heat flux as
  $k_{y, \min}\rho_i\to 0$ is not carried by the ever smaller minimum
  wavenumbers but by the contributions of modes $0.07\le k_y\rho_i \le 0.35$
  in the vicinity of the peak (contributing to at least $90\%$ of the heat
  flux), range fully covered by all simulations, except for the largest
  considered $k_{y, \min}\rho_i=0.14$.  Note that the inertial range
  ($k_y\rho_i\gtrsim 0.35$) remains essentially identical over all runs. 
  
  %The increasing heat flux as $k_{y, \min}\rho_i$ is reduced thus takes place in conjunction with decreasing zonal $E\times B$ shearing rates, which is consistent with the established paradigm that shearing of turbulent eddies by zonal flows is the dominant saturation mechanism for ITG-driven turbulence.
  {\it A priori}, a straightforward explanation for the decreasing
  $E\times B$ shearing rate leading to increased heat fluxes as
  $k_{y, \min}\rho_i$ decreases, is the reduction in the radial density of stationary
  shear layers at LMRSs. Let us indeed recall that the distance between LMRSs
  is given by $\Delta x_{\rm LMRS} = 1/(k_{y, \min}\hat{s})$. Note as well the
  remarkable fact that the radial width and amplitude of the shear layers at
  LMRSs remains essentially invariant when varying $k_{y, \min}\rho_i$, as
  illustrated in figure~\ref{omegaecrossbionsvsx_sepspcompare}(b).  The
  contribution to the total shearing rate estimate (\ref{RMS_xt(w_eff)}) from
  the stationary component of the shearing rate profile $\langle\omega_{\rm
    eff}\rangle_t(x)$ can be calculated by:
  \begin{equation}
    \label{RMS_x(<w_eff>_t)}
    {\rm RMS}_{x}(\langle\omega_{\rm eff}\rangle_t)
    =
    \left[\,
    \langle\,
    \left(\,
        \langle\omega_{\rm eff}\rangle_t
    \,\right)^2
    \,\rangle_{x}
    \,\right]^{1/2},
  \end{equation}
  and has been plotted in log-log scale as a function of $k_{y, \min}\rho_i$
  in figure~\ref{omegaeffvskymin}(b). As expected, this system average of the
  stationary shearing rate profile decreases algebraically for kinetic
  electron simulations with decreasing $k_{y, \min}$, while for the adiabatic
  electron simulations there is no obvious dependence on $k_{y, \min}\rho_i$
  as the mechanism developing the prominent fine stationary structures on
  $\langle\omega_{\rm eff}\rangle_t(x)$ is absent in this case. 
  %The finite values of $ {\rm RMS}_{x}(\langle\omega_{\rm eff}\rangle_t)$ appearing in figure~\ref{omegaeffvskymin}(b) for the adiabatic electron cases are in fact essentially the result of statistical error due to finite simulation time  $t_{\rm sim}$. 
  
\begin{figure}
\centering
\includegraphics[width=0.45\columnwidth] {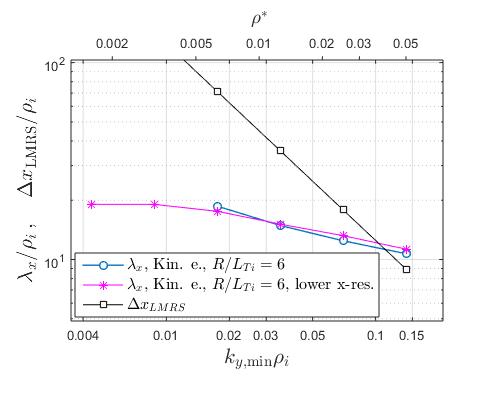}
  \caption{Blue circles denote the radial correlation length
    $\lambda_x$ of turbulent eddies in units of $\rho_i$ as a function
    of $k_{y, \min}\rho_i$ for kinetic electron simulations and
    $R/L_{T,i}=6$ (all parameters as given in
    table~\ref{Parameterset}). Magenta asterisks correspond to results
    obtained with the same parameters except for a decreased radial
    resolution (by a factor $4$), which enabled to carry out runs with
    $k_{y, \min}\rho_i$ values as low as $\sim 5\cdot 10^{-3}$ (these
    simulations are discussed in more detail in relation with
    Fig. \ref{Qivskymin_4_dxb4}). Distance $\Delta x_{LMRS}=1/(k_{y,
      \min}\hat{s})$ between LMRSs is plotted with black squares.}
  \label{Corrvskymin}
\end{figure}

  This explanation for the increase of turbulent fluxes as a result of
  the decrease in the radial density of stationary zonal $E\times B$
  shear layers is however not satisfactory at closer scrutiny. This is
  made clear by the results presented in figure~\ref{Corrvskymin},
  where the radial correlation length of turbulent eddies $\lambda_x$
  is plotted as a function of $k_{y, \min}\rho_i$ for the case with
  kinetic electrons and $R/L_{Ti}=6$.  The radial correlation length
  is defined as $\lambda_x=\langle\lambda_x(y)\rangle_y$, with
  $\lambda_x(y)$ estimated as that smallest value of $\Delta x$ for
  which the auto-correlation function $R(\Delta x,y)=\int
  \Phi'^*(x-\Delta x,y)\Phi' (x,y)dx$ is $1/e$ times its maximum value
  (this maximum is reached for $\Delta x = 0$ and $e$ is the base of
  the natural logarithm).  $\Phi'(x,y)$ is the scalar potential
  evaluated at the outboard midplane, with the zonal ($k_y=0$)
  component removed:
  $\Phi'(x,y)=\Phi(x,y,z=0)-\langle\Phi(x,y,z=0)\rangle_y$. This
  radial correlation length increases as $k_{y, \min}\rho_i$
  decreases, but with a much weaker scaling (fit provides
  $\lambda_x/\rho_i \sim k_{y, \min}\rho_i^{-0.27}$ over the
  considered range $k_{y, \min}\rho_i = 10^{-2} - 10^{-1}$) than the
  increase of the distance $\Delta x_{\rm LMRS}/\rho_i\sim(k_{y,
    \min}\rho_i)^{-1}$ between the main stationary shear layers
  located at LMRSs. Below a sufficiently small value of $k_{y,
    \min}\rho_i$, one thus clearly has $\lambda_x \ll \Delta x_{\rm
    LMRS}$. That is, the turbulent eddies are getting actively sheared
  and broken {\it in between} low order MRSs where the stationary
  zonal shear flows are insignificant. The stationary shear layers are
  therefore not expected to play a major role in the saturation of
  turbulence as $k_{y, \min}\rho_i\to 0$. We therefore conclude that
  as $k_{y, \min}\rho_i\to 0$, the saturation of turbulence through
  the break-up of turbulent eddies is mainly to be attributed to the
  {\it fluctuating} component of zonal flows. This is discussed in
  detail in the following.
  
  The time dependent component (as opposed to the stationary
  component) of the $E\times B$ zonal flow and associated shearing
  rate thus appears to control the saturation of turbulence and
  related flux levels {\it between} LMRSs. An estimate for the
  amplitude of this fluctuating part of the shearing rate is provided
  by computing the radial profile of the Standard Deviation ${\rm
    SD}_t(\omega_{\rm eff})$ of $\omega_{\rm eff}$ around the
  stationary component $\langle\omega_{\rm eff}\rangle_t$:
  \[
  {\rm SD}_{t}(\omega_{\rm eff})(x)
  = 
  \left[\, \langle\, \left( \omega_{\rm eff}
    - \langle\omega_{\rm eff}\rangle_t \right)^2 \,\rangle_{t}
    \,\right]^{1/2}
  \]
  The radial dependence of ${\rm SD}_{t}(\omega_{\rm eff})$ has been
  added to figure~\ref{omegaecrossbionsvsx_sepspcompare}(b) for the
  cases with kinetic electrons, $R/L_{T_i}=6$ and both $k_{y,
    \min}\rho_i = 0.035$ and $k_{y, \min}\rho_i = 0.0175$. Based on
  figure~\ref{omegaecrossbionsvsx_sepspcompare}(b), it appears that
  the fluctuating part of the shearing rate remains essentially
  constant across the radial extent of the system. Furthermore, its
  amplitude for the considered values of $k_{y, \min}\rho_i$ still
  remains significant, {\it i.e.}  larger then $\gamma_{\max}$, and of
  the same order as the maximum amplitude of the time-averaged
  $\langle\omega_{E\times B}\rangle_t$ structures. A decrease of the
  fluctuating component with decreasing $k_{y, \min}\rho_i$ is also
  observed. This is summarized in figure~\ref{omegaeffvskymin}(c),
  where the radial average of the fluctuating component, estimated by
  \begin{equation}
    \label{SD_xt(w_eff)}
          {\rm SD}_{x,t}(\omega_{\rm eff})
          =
          \left[\,
            \langle\,
            \left(
            \omega_{\rm eff} - \langle\omega_{\rm eff}\rangle_t
            \right)^2
    \,\rangle_{x,t}
    \,\right]^{1/2},
  \end{equation}
  has been plotted. While ${\rm SD}_{x,t}(\omega_{\rm eff})$ decreases
  with $k_{y, \min}\rho_i$ in both adiabatic and kinetic electron
  simulations, the scaling is much stronger in the latter case. For
  instance, in the considered range of $k_{y, \min}\rho_i$, ${\rm
    SD}_{x,t}(\omega_{\rm eff})\sim(k_{y, \min}\rho_i)^\alpha$, where
  $\alpha=0.08$ for the case with adiabatic electrons at $R/L_{Ti}=15$
  while $\alpha=0.32$ for the case with kinetic electrons at
  $R/L_{Ti}=6$.
    
    Note that according to the definitions
  (\ref{RMS_xt(w_eff)}), (\ref{RMS_x(<w_eff>_t)}) and (\ref{SD_xt(w_eff)}),
  one has
  \[
  \left[
    {\rm RMS}_{x,t}(\omega_{\rm eff})
    \right]^2 
  = 
  \left[
    {\rm RMS}_{x}(\langle\omega_{\rm eff}\rangle_t)
    \right]^2 
  + 
  \left[
    {\rm SD}_{x,t}(\omega_{\rm eff})
    \right]^2.
  \]
  The decrease of both the fluctuation level of the shearing rate,
  shown in figure~\ref{omegaeffvskymin}(c), along with the decrease in
  the density of stationary shearing layers at LMRSs, reflected by
  figure~\ref{omegaeffvskymin}(b), thus provides a more complete
  picture of the decrease in the total system average of the shearing
  rate $\omega_{\rm eff}$ as $k_{y,\min}\rho_i$ decreases, shown in
  figure~\ref{omegaeffvskymin}(a). While the reason for the decrease
  in the contribution to the shearing rate from stationary zonal flows
  has already been discussed and is relatively obvious, the reason for
  the decrease of the contribution from fluctuating zonal flows is
  not. To explain the latter, it is necessary to understand and
  analyze the different mechanisms driving the zonal flows. This is
  done in the following section.

%
%----------------------------------------------------------------------
%
\section{Analyzing zonal flow drive}\label{AnalyzingZFdrive}
%
%------------------------------
%
 In this section, the drive of zonal flows via the two
  main mechanisms, the modulational instability and self-interaction,
  are studied. In particular, the statistical properties of these
  different zonal flow drives are analysed using various diagnostics
  techniques, with the primary objective of understanding why the
  fluctuating zonal flow levels decrease with decreasing
  $k_{y,\min}\rho_i$ as seen in figure~\ref{omegaeffvskymin}(c).

In order to ensure a systematic study, this section has been organised
as follows. To start, it is essential to identify the physical
quantity(ies) representing the drive of zonal flows ($\sim$ modes
$k_{y}\rho_i=0$) from turbulence ($\sim$ modes $k_{y}\rho_i\neq
0$). In \S~\ref{ZFproxy}, it is found that Reynolds stress can
be used as a convenient proxy for quantifying the drive of zonal
flows. The two main mechanisms driving zonal flows are then discussed
in the following two subsections: a summary of the well known
modulational instability mechanism is given in
\S~\ref{Refslab}, followed by a detailed description of the
self-interaction mechanism in \S~\ref{Reftoroidal}. These two
mechanisms and the nature of their drives are illustrated first in a
simple nonlinear set-up, referred to as the ``reduced simulations'',
in \S~\ref{Refreduced}.
%Here, through a comparison between the two different electron models, the adiabatic and kinetic models, it is found that modulational instability is the dominant zonal flow driving mechanism with adiabatic electrons, while self-interaction dominates in kinetic electron simulations. 
This is followed by providing the evidence for self-interaction and
modulational instability in full turbulence simulations, discussed in \S~\ref{Sec. SI in turbulent sims} and \S~\ref{Sec. MI in
  turbulent sims} respectively.
%In these sections, the dominance of modulational instability and self-interaction in adiabatic and kinetic electron simulations respectively is found to hold for turbulence simulations as well. 
In the latter subsection, using a bicoherence-like analysis and an
estimate of the correlation between the different $k_y$ contributions
to Reynolds stress, it is shown that the drive of zonal flows via
self-interaction from each $k_y$ are essentially incoherent and
decorrelated, unlike that via modulational instability. This result
will then be used in the following \S~\ref{Statdep}, to show how
such a decorrelated drive could explain the observed decrease in
fluctuating zonal flow levels with decreasing $k_{y,\min}\rho_i$.

\subsection{Reynolds stress as a proxy for the drive of zonal flows}\label{ZFproxy}

Zonal flows are linearly stable and are driven by turbulence through the
  quadratic E$\times $B non-linearity appearing in the gyrokinetic
  equation. To better understand their evolution, we will analyze the
  properties of Reynolds stress \citep{Diamond2005}, more exactly the
  off-diagonal component $\langle \tilde{V}_x\tilde{V}_\chi\rangle$ of the
  Reynolds stress tensor resulting from the combination of fluctuating
  $E\times B$ flow components in the radial and poloidal directions. We will
  justify in the following that this Reynolds stress can be considered
  as a valid proxy of the zonal flow drive \citep{Weikl2018}
  . By analyzing Reynolds stress we will be able to identify
  the different possible mechanisms driving zonal flows, their statistical properties,
  and relative importance.
  
  We start by considering an approximate evolution equation for the
  shearing rate $\omega_{E\times B}$ associated to the $E\times B$ zonal
  flows. As shown in references \citep{Parra2009,Abiteboul2011,AbiteboulPhD}, such an
  equation can be obtained from the radial conservation equation for the total
  gyrocenter charge density, which in turn is derived by taking the
  appropriate velocity moment and flux-surface -average of the gyrokinetic
  equation. The approach in reference \citep{,AbiteboulPhD} has been considered here, but starting from
  the gyrokinetic equation in the limit of the local (flux-tube) delta-f model rather
  than the global full-f model considered by \citet{,AbiteboulPhD}
  . Furthermore, assuming the electrostatic
  limit, invoking $m_e/m_i\ll 1$, and making use of the quasi-neutrality
  equation in the long wavelength approximation (final result thus valid only
  to second order in $k_\perp\rho_i$), leads to a relation that can be
  interpreted as a generalized vorticity equation:
  \begin{equation}
    \label{Eq_shearRS}
%    \frac{\partial }{\partial t}\Omega+\frac{\partial }{\partial t}\Pi
    \frac{\partial }{\partial t}
    (\Omega+\Pi)
    = 
%    \frac{\partial^2}{\partial x^2}\mathcal{R} 
%    + 
%    \frac{\partial^2}{\partial x^2}\mathcal{P}
    \frac{\partial^2}{\partial x^2}
    (\mathcal{R} + \mathcal{P})
    + 
    \frac{\partial}{\partial x}\mathcal{N}.        
  \end{equation}
  One identifies on the left hand side of equation~(\ref{Eq_shearRS}) the
  generalized vorticity term $\Omega+\Pi$, composed of the actual vorticity
  associated to zonal flows and closely related to the zonal flow shearing
  rate $\omega_{E\times B}$ (by neglecting the $z$ dependence of $B_0$
  and $g^{xx}\simeq 1$):
  \[
  \Omega
  =
  n_{0,i}m_i
  \frac{\partial^2}{\partial x^2}
  \left\langle
  \frac{g^{xx}}{B_0^2}\,\Phi
  \right\rangle_{yz}
  \simeq
  \frac{n_{0, i}m_i}{B_0}\,\omega_{E\times B},
  \]
  as well as a perpendicular pressure term, related to lowest order finite ion
  Larmor radius effects:
  \[
  \Pi
  =\frac{m_i}{2q_i}
  \frac{\partial^2}{\partial x^2}
  \left
  \langle
  \frac{g^{xx}}{B_0^2}\,
  P_{\perp,i}
  \right
  \rangle_{yz},
  \]
  with the fluctuating part of the perpendicular ion gyrocenter pressure
  $P_{\perp,i}$ expressed in terms of the corresponding gyrocenter
  distribution fluctuation $\delta f_i$:
  \[
  P_{\perp,i} = \int \mu B_0\,\delta f_i\ d^3v.
  \]
  On the right hand side of equation~(\ref{Eq_shearRS}) one identifies the second
  radial derivative of a term $\mathcal{R}=(n_{0,i}m_i/\mathcal{C})\,{\rm RS}$,
  proportional to the Reynolds stress component ${\rm RS}$ driving zonal
  flows:
  \begin{equation}
    \label{EqRSdef}
    {\rm RS}
    =
    \left\langle
    \frac{1}{B_0^2}\frac{\partial \Phi}{\partial y}
    \left(
    g^{xx}\frac{\partial \Phi}{\partial x}
    +
    g^{xy}\frac{\partial \Phi}{\partial y}
    \right)
    \right\rangle_{yz}.
  \end{equation}
   This Reynolds stress term also has a finite Larmor radius
  correction term $\mathcal{P}$, again expressed in terms of $P_{\perp,i}$:
  \[
  \mathcal{P}
  =
  \frac{m_i}{2q_i\,\mathcal{C}}
  \left\langle
  \frac{1}{B_0^2}
  \left(
  g^{xx}\,
  \frac{\partial\Phi}{\partial x}
  \frac{\partial P_{\perp,i}}{\partial y}
  +
  2\,g^{xy}\,
  \frac{\partial \Phi}{\partial y}
  \frac{\partial P_{\perp,i}}{\partial y}
  +
  g^{xx}\,
  \frac{\partial \Phi}{\partial y}
  \frac{\partial P_{\perp,i}}{\partial x}
  \right)
  \right\rangle_{yz}.
  \]
  The last contribution on the right hand side of equation~(\ref{Eq_shearRS}) is
  the radial derivative of the so-called neoclassical term $\mathcal{N}$,
  related to both curvature and $\nabla B$ drifts:
  \[
  \mathcal{N}
  =
  -\sum_{\rm species}
  \frac{2\upi\,\mathcal{C}}{m}
  \left\langle 
  \gamma_2
  \frac{\partial B_0}{\partial z}
  \int dv_\parallel d\mu\,
  \frac{mv_\parallel^2 + \mu B_0}{B_0^2}
  \left(
  \delta f + \frac{q\,\bar{\Phi}}{T_0} f_0
  \right) 
  \right\rangle_{yz},
  \]
  with $f_0$ the Maxwellian background distribution,
  $\gamma_2{=}g^{xx}g^{yz}-g^{xy}g^{xz}$ and the constant
  $\mathcal{C}{=}B_0/|\nabla x\times\nabla y|$.

    The different terms appearing in (\ref{Eq_shearRS}) can be monitored
  as a diagnostic along a gyrokinetic simulation. For the purpose of
  verification, this diagnostic was first applied to simple test cases,
  including the Rosenbluth-Hinton problem \citep{Rosenbluth1998} addressing the
  linear dynamics of zonal fluctuations [in this case only the linear
    neoclassical term contributes on the RHS of (\ref{Eq_shearRS})], as well
  as the non-linear decay of an initially single unstable eigenmode (as
  discussed in \S\ref{Refreduced}). For these simple tests, only longer
  wavelength modes were considered, so as to stay well within the limit
  $k_\perp\rho_i\ll 1$ assumed for the derivation of relation
  (\ref{Eq_shearRS}).

  After this successful initial verification phase, the terms appearing in
  equation~(\ref{Eq_shearRS}) were monitored and compared for the fully developed
  turbulence simulations studied in this paper, considering the reference
  gradient case $R/L_{T,i}=6$. In order to validate Reynolds stress as a good
  proxy for the drive of zonal flows, the correlation between
  $\Omega\sim\omega_{E\times B}$ and the two non-linear driving terms
  $\partial^2\mathcal{R}/\partial x^2$ and $\partial^2\mathcal{P}/\partial
  x^2$ in (\ref{Eq_shearRS}) was estimated over these simulations. To this
  end, the following correlation estimator between two observables $a$ and
  $b$, functions of the radial variable $x$ and time $t$, was applied:
  \[
    {\rm Corr}(a,b)
    =
    \frac{\sigma_{x,t}(a, b)}{\sigma_{x,t}(a)\,\sigma_{x,t}(b)}
    =
    \frac{
      \langle\,
      (a-\langle a\rangle_{x,t})\,
      (b-\langle b\rangle_{x,t})
      \,\rangle_{x,t}}
    {\sqrt{\langle\,(a-\langle a\rangle_{x,t})^2\,\rangle_{x,t}}\,
     \sqrt{\langle\,(b-\langle b\rangle_{x,t})^2\,\rangle_{x,t}}}.
  \]
  Significant correlation estimates were obtained in this way between $\Omega$
  and the Reynolds stress drive term $\partial^2\mathcal{R}/\partial
  x^2$. Correlation values were found to increase further by considering only the
  longer wavelength contributions, achieved by filtering out $|k_x|\rho_i >
  0.5$ and $|k_y|\rho_i>0.5$ Fourier modes from the signals. This is in
  agreement with the long wavelength approximation assumed for deriving
  equation~(\ref{Eq_shearRS}). The positive correlation values ${\rm Corr} = 0.77$
  and $0.37$ were obtained in this way between $\Omega$ and
  $\partial^2\mathcal{R}/\partial x^2$ for the adiabatic and kinetic electron
  turbulence simulations respectively, while the correlation between $\Omega$
  and $\partial^2\mathcal{P}/\partial x^2$ provided the values ${\rm Corr} =
  0.75$ and $0.40$ respectively. These results validate considering the
  Reynolds stress term $\partial^2{\rm RS}/\partial x^2$ as a proxy for the
  drive of zonal flow shear $\omega_{E\times B}$.
  
  In the following, it will be insightful to consider the contributions from
  different $k_y$ Fourier modes components of the fluctuating fields to the
  Reynolds stress term ${\rm RS}$. Relation (\ref{EqRSdef}) for ${\rm RS}$ can
  indeed be written as a sum over $k_y$:
  \begin{equation}
    \label{EqRS, sum}
    {\rm RS}(x, t) 
    =
    \sum_{k_y > 0} \hat{\rm RS}_{k_y}(x, t),
  \end{equation}
  with the contribution from the $k_y$ Fourier mode $\hat\Phi_{k_y}(x, z, t) =
  \frac{1}{L_y} \int_0^{L_y} \Phi\exp(-{\mathrm i}k_yy)\,dy$ given by
  \begin{equation}
    \label{RS_ky}
    \hat{\rm RS}_{k_y}(x, t)
    =
    2\,\text{Re}
    \left[\;
      \left\langle
      \frac{1}{B_0^2}\,k_y\hat\Phi_{k_y} 
      \left(
      g^{xx}{\rm i}\,\frac{\partial \hat\Phi_{k_y}^\star}{\partial x}
      +
      g^{xy} k_y \hat\Phi_{k_y}^\star
      \right)
      \right\rangle_z
    \right],
  \end{equation}
  having invoked the reality condition
  $\hat\Phi_{-k_y}=\hat\Phi_{k_y}^\star$. Considering as well the $k_x$
  Fourier mode decomposition of $\Phi$, each of these $k_y$ contributions can
  also be written as follows:
  \begin{equation}
    \label{RS_ky in kx-repr}
    \hat{\rm RS}_{k_y}(x, t)
    =
    2\,\text{Re}
    \{\;
    \sum_{k_x, k_x''}
    \left\langle
    \frac{1}{B_0^2}\,k_y
    \left(
    g^{xx} k_x''
    +
    g^{xy} k_y 
    \right)\hat\Phi_{k_x,k_y} \hat\Phi_{k_x'',k_y}^\star
    \right\rangle_z
    \exp[{\rm i}(k_x-k_x'')x]
    \;\},
  \end{equation}
  illustrating the drive of zonal modes $(k_x'=k_x-k_x'', 0)$ through
  non-linear interaction between Fourier modes $(k_x, k_y)$ and $(k_x'',
  k_y)$, extensively discussed in sections \S\ref{Refslab} and \S\ref{Reftoroidal}.

  For the study carried out in this paper, it is essential to understand how
  different drift modes may non-linearly interact to drive
  zonal flows via Reynolds stress. To start, we recall in the next subsection
  the basic mechanism underlying the drive of zonal flows in a simple
  shearless slab system before considering the more complex case of direct
  interest to us, {\it i.e.} of a sheared toroidal system.
  
%
%------------------------------
%
\subsection{Modulational instability in shearless slab system}
\label{Refslab}
The drive of zonal flows by microturbulence has been extensively
studied in the literature considering a simple slab-like plasma
confined by a uniform, shearless magnetic field. Such a system was in
particular addressed in the original work by \citet{Hasegawa1978}, where a cold fluid model was assumed for ions and an
adiabatic response for electrons. In this model, choosing an orthogonal
Cartesian coordinate system $(x, y, z)$, the magnetic field is aligned along
$z$, $\mathbf B = B\mathbf e_z$ and the background density inhomogeneities
along $x$, $\nabla n_0 = ({\rm d} n_0/{\rm d}x)\mathbf e_x$. The corresponding well-known
model equation for the non-linear evolution of the electrostatic potential
$\Phi$ associated to the fluctuating fields describes the essentially
two-dimensional drift wave turbulence in the $(x, y)$ plane perpendicular to
the magnetic field. This model equation led to a first understanding of the
generation of zonal flows along $y$, {\it i.e.} in the direction both
perpendicular to $\mathbf B$ and the direction of inhomogeneity. The emergence
of such large scale flows can in particular be explained as the result of an
anisotropic inverse cascade of energy related to the conservation of energy
and enstrophy in the 2-dimensional turbulence. The emergence of zonal
flows can also be understood at the level of elementary non-linear processes
as recalled in the following.
  
In a shearless slab system, the spatial dependence of linear eigenmodes are
given by a {\it single} Fourier mode $\Phi(x,
y)\sim\Phi_\bk\exp({\rm i}\bk\cdot\bx)$, with $\bx = x\,\be_x+y\,\be_y$ and $\bk =
k_x\be_x+k_y\be_y$. The corresponding time dependence is of the form
\mbox{$\sim \exp(-{\rm i}\omega_\bk t)$}, with $\omega_\bk$ the eigenfrequency of
the mode. In the simple Hasegawa-Mima model, one has $\omega_\bk =
k_y\,v_d/(1+k^2)$, with $\bv_d = -(T_e/eB)({\rm d}\log n_0/{\rm d}x)\, \be_y$ these
eigenmodes results from the {\it quadratic} non-linearity in the Hasegawa-Mima
equation, related to the $\bv_{E\times B} = (-\nabla\Phi\times\bB)/B^2$
drift. The elementary non-linear interaction thus involves a {\it triplet} of
Fourier modes $\bk$, $\bk'$, $\bk''$ verifying the wave vector matching condition $\bk =
\bk'+\bk''$, where each of the modes, {\it e.g.} $\bk$, is coupled to the two
others, $\bk'$ and $\bk''$ in this case. In case of frequency matching
$\omega_\bk\simeq\omega_{\bk'}+\omega_{\bk''}$ and under the condition
$k'<k<k''$, one can have a resonant decay of mode $\bk$, {\it i.e.} a transfer
of energy from this mode, into the daughter modes $\bk'$ and $\bk''$. This
basic process is referred to as the resonant 3-wave interaction mechanism. One
can furthermore show that in the case where mode $\bk$ represents a drift
wave, {\it i.e.}  typically with $|k_y| \gg |k_x|$, the decay happens
preferentially (meaning with a higher growth rate of decay) if one of the
daughter wave vectors, {\it e.g.} $\bk'$, is (nearly-) aligned along the $x$
direction, $|k_x'| \gg |k_y'|$ \citep{Hasegawa1979}. The $E\times B$ flow
associated to the daughter mode $\bk'$ is then obviously along $y$, thus
explaining the emergence of zonal flows in this direction. Such a mode with
vector aligned along the direction of inhomogeneity $\be_x$ is thus referred
to as a zonal mode.
  
Let us still further consider the decay of a pump drift wave $\bk = (k_x,
k_y)$, $k_y\neq0$, into a zonal mode $\bk'= (k_x', 0)$ and the second daughter
wave $\bk'' = \bk-\bk' = (k_x-k_x', k_y)$, itself a drift wave. The non-linear
interaction between the two drift waves $\bk$ and $\bk''$ thus provides the
drive to the zonal mode $\bk'$ via the Reynolds stress ${\rm RS}$ discussed in
\S\ref{ZFproxy}, while the non-linear coupling between the original drift
wave $\bk$ and zonal mode $\bk'$, leading to the growth of the daughter drift
wave $\bk''$, actually represents the {\it shearing} of the drift mode $\bk$
by the zonal flows associated to $\bk'$.
  
Variations to the simple Hasegawa-Mima model have been considered in the
literature. In particular, the Enhanced Hasegawa-Mima model
\citep{Krommes2000,Gallagher2012} accounts for the fact that the adiabatic
electron response is inhibited for magnetic surface-averaged fluctuations,
{\it i.e.} modes $\bk$ with $k_y=0$, in other words zonal modes, which leads
them to having a reduced effective inertia compared to standard drift waves
with $k_y\neq 0$. This effect results in an amplification of the decay rates
of drift waves into zonal modes and thus to an enhancement of corresponding
energy transfer.

A further refinement to the basic driving mechanism of zonal flows is obtained
by accounting for the fact that given an initial large amplitude drift mode
$\mathbf k$, decaying into a zonal mode $\mathbf k'$, both triplet
interactions $[\mathbf k, \mathbf k', \mathbf k'' = \mathbf k-\mathbf k']$ and
$[\mathbf k, -\mathbf k', \mathbf k''' = \mathbf k+\mathbf k']$ may be
simultaneously resonant, {\it i.e.}  $\omega_{\mathbf k}\simeq\omega_{\mathbf
  k'}+\omega_{\mathbf k''}$ and $\omega_{\mathbf k}\simeq-\omega_{\mathbf
  k'}+\omega_{\mathbf k'''}$. This coupled pair of 3-wave interactions leads
to an effective 4-wave interaction involving modes $\mathbf k$, $\mathbf k'$,
$\mathbf k''$ and $\mathbf k'''$, referred to as the modulational instability
mechanism \citep{Gallagher2012}.
  
An important point to emphasize is that, in both the case of the simple
resonant 3-wave interaction or the more general modulational instability,
resonant coupling between the initial large amplitude (pump) drift mode
$\bk=(k_x, k_y\neq 0)$ and a zonal mode $\bk'=(k_x', k_y'=0)$ is established
via either one sideband $\bk''=\bk-\bk'$, or respectively two sidebands $\bk''$
and $\bk'''=\bk+\bk'$, where all these Fourier modes are {\it linearly
  decoupled} from each other. As a result, in the situation where in addition
to the pump drift mode $\bk$, the zonal mode $\bk'$ itself already has a finite
initial amplitude and a well defined phase, the other daughter waves
$\bk''$ (and $\bk'''$) are {\it free} to adapt their phases to ensure
a resonant interaction and thus an efficient energy transfer from the pump to
the zonal mode, resulting in the further amplification of this zonal mode. In
fact, a finite amplitude zonal mode can {\it stimulate} the decay of
multiple non-zonal modes, thus leading to {\it coherent} (as a result of the frequency matching involved) and therefore {\it
  correlated} contributions from these non-zonal modes to the Reynolds stress
drive of the zonal mode.

Under realistic conditions, multiple elementary mechanisms such as the ones
discussed above (resonant 3-wave interaction / modulational instability)
happen both simultaneously and successively in time, leading to an expanding
set of Fourier modes, and ultimately to a fully developed turbulent spectrum.
  
%
%------------------------------
%  
\subsection{Self-interaction mechanism in sheared toroidal system}
\label{Reftoroidal}

As in the shearless slab geometry, in the case of a tokamak system, {\it
  i.e. } based on a sheared axisymmetric toroidal magnetic geometry, which is
of main interest to our study, the modulational instability involving resonant
3-wave interactions remains an essential driving mechanism of zonal
flows \citep{Chen2000}. In a tokamak however, one must distinguish another form
of the non-linear interaction leading to the drive of zonal modes. This
mechanism, referred to as self-interaction, is specific to systems presenting
magnetic shear and is explained in the following.

As already discussed in \S\ref{secGENE}, parallel boundary conditions lead
to the linear coupling of $k_x$ Fourier modes. In particular, according to
equation~(\ref{linear eigenmode, Fourier rep.}), the electrostatic potential field
$\Phi_{\rm L}$ of a linear microinstability eigenmode with fixed $k_{x0}$ and
$k_y\neq 0$ is composed of Fourier modes $\hat{\Phi}_{k_{x0} + p\,2\upi
  k_y\hat{s},\,k_y}$, $p\in\mathbb{Z}$. The spatial dependence of the
corresponding eigenmode structure is thus given by:
\begin{equation}
  \label{linear eigenmode, phi}
  \Phi_{\rm L}(x, y, z) 
  = 
  \tilde{\Phi}_{k_{x0},k_y}(x,z)\exp({\rm i}k_yy) + {\rm c.c.},
\end{equation}
where ${\rm c.c.}$ stands for the complex conjugate [considered here to ensure
  that $\Phi_L$ is real-valued, essential for computing the quasi-linear
  estimate in (\ref{EqRSqlin})], and with the complex-valued $(x,
z)$-dependent envelope given by:
\begin{equation}
  \label{linear eigenmode, phikx0ky}
  \tilde{\Phi}_{k_{x0},k_y}(x,z)
  =
  \sum_{p=-\infty}^{+\infty}
  \hat{\Phi}_{k_{x0} + p2\upi k_y\hat{s},\,k_y}(z) 
  \exp[{\rm i}(k_{x0} + p\,2\upi k_y\hat{s})x].
\end{equation}
These linearly coupled $k_x$ Fourier modes, all having same $k_y$, drive the
zonal modes $(k_x', 0)$, with $k_x'=p'\,2\upi k_y\hat{s}$, $p'\in\mathbb{Z}$,
forming a set of harmonics. Indeed, any two Fourier modes $\hat{\Phi}_{k_{x0}
  + p\,2\upi k_y\hat{s}, \,k_y}$ and $\hat{\Phi}_{k_{x0} + p''\,2\upi
  k_y\hat{s},\,k_y}$ composing the physical eigenmode will drive, via three
Fourier mode coupling, the zonal mode $\hat{\Phi}_{p'\,2\upi k_y\hat{s},\,0}$
with $p'=p-p''$ [see equation~(\ref{RS_ky in kx-repr})]. Note that this drive of
zonal modes is via the same quadratic non-linearity in the gyrokinetic
equation related to $E\times B$ drifts as the one driving zonal modes through
the modulational instability, but in this case involving Fourier modes which
are already {\it linearly} coupled to each other.

Assuming that the relative phases between the Fourier modes $\hat{\Phi}_{k_{x,
    0} + p\,2\upi k_y\hat{s},\,k_y}$ remains set by the linear coupling, even
during the non-linear turbulent evolution (to what extent this assumption
holds is validated in \S\ref{Sec. SI in turbulent sims}), the phases of the
associated contributions to Reynolds stress driving the zonal modes is {\it
  fixed}. This translates in direct space to an essentially fixed
periodic radial dependence (with period $\Delta x_{\rm MRS} = 1/k_y\hat{s}$
corresponding to the distance between MRSs) of the contribution to Reynolds
stress through this self-interaction mechanism from a given $k_y$ eigenmode,
the overall magnitude of this contribution obviously varying in time as the
amplitude of the eigenmode evolves.  This is a critical difference compared to
the drive of zonal modes $\bk'$ through the modulational instability discussed
in \S\ref{Refslab}, where the relative phases between the pump mode $\bk$
and sidebands $\bk''$, $\bk'''$ are {\it free} to adapt to enable a coherent
({\it i.e.} in phase) drive of a given zonal mode.

\begin{figure}
\centering
\includegraphics[width=0.65\columnwidth]{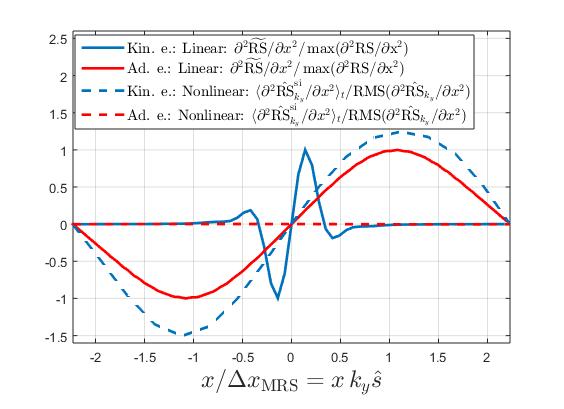}
\caption{Solid lines indicate the quasi-linear estimate of
  $\partial^2\widetilde{\rm RS}_{k_{x0},k_y}/\partial x^2$ normalised to its
  maximum value as a function of $x$ for the linear eigenmode with
  $k_{x0}=0$, $k_y\rho_i=0.28$ and either kinetic (blue) or adiabatic (red)
  electrons. Dashed lines denote the time-average
  $\langle\partial^2\hat{\rm RS}^{\rm si}_{k_y}/\partial x^2\rangle_t$ 
  normalised with respect to the RMS in time of the total contribution
  $\partial^2\hat{\rm RS}_{k_y}/\partial x^2$, for $k_y\rho_i=0.28$,
  in turbulence simulations considering
  either kinetic (blue) or adiabatic (red) electrons. $k_{y,\min}\rho_i=0.035$, $R/L_{T_i}=6$ and other parameters are given in table~\ref{Parameterset}.}
\label{Lind2RS}
\end{figure}

Figure~\ref{Lind2RS} plots the quasi-linear estimates of this self-interaction
contribution to the Reynolds stress drive term $\partial^2{\rm RS}/\partial
x^2$ from the linear eigenmode with $k_{x0}=0$, $k_{y}\rho_i = 0.28$, and
$R/L_{T_i} =6$, whose ballooning structures are given in
figure~\ref{Ballooning}, considering both the case of kinetic and adiabatic
electron response. These results are obtained by inserting the corresponding
eigenmode structure (\ref{linear eigenmode, phi}) into relations (\ref{RS_ky})
and (\ref{RS_ky in kx-repr}), leading to the Reynolds stress contribution
denoted $\widetilde{\rm RS}_{k_{x0},k_y}(x)$:
\begin{flalign}
  \nonumber
  \widetilde{\rm RS}_{k_{x0},k_y}(x) 
  & = 
  {\rm RS}[\Phi_L]
  =
  \hat{\rm RS}_{k_y}[\tilde\Phi_{k_{x0}, k_y}] \\
  \nonumber
  & =
  2\,\text{Re}
  \left[\;
    \left\langle
    \frac{1}{B_0^2}\,k_y\tilde\Phi_{k_{x0}, k_y} 
    \left(
    g^{xx}{\rm i}\,\frac{\partial \tilde\Phi_{k_{x0}, k_y}^\star}{\partial x}
    +
    g^{xy} k_y \tilde\Phi_{k_{x0}, k_y}^\star
    \right)
    \right\rangle_z
    \;\right]\\
    \label{EqRSqlin}
    & =
    2\,\text{Re}
    \left[\;
    \sum_{p'=-\infty}^{+\infty}     
%    \exp[ip'2\upi k_y\hat{s}x]
    e^{{\rm i}p'2\upi k_y\hat{s}x}
    \sum_{p=-\infty}^{+\infty}
    \left\langle
    \frac{1}{B_0^2}\,k_y
    \left(
    g^{xx} k_x''
    +
    g^{xy} k_y 
    \right)\hat\Phi_{k_x,k_y} \hat\Phi_{k_x'',k_y}^\star
    \right\rangle_z
    \;\right].
\end{flalign}
In the last equality of relation (\ref{EqRSqlin}), $k_x = k_{x0} + p\,2\upi
k_y\hat{s}$, $k_x'' = k_{x0} + p''\,2\upi k_y\hat{s}$ and $p'' = p -
p'$. This relation also clearly points out how this contribution to Reynolds
stress from a given $k_y$ eigenmode through self-interaction is periodic with
period $\Delta x_{\rm MRS} = 1/k_y\hat{s}$.

The overall amplitude of the quasi-linear estimates shown in figure~\ref{Lind2RS}
are naturally irrelevant. Furthermore, as these contributions to Reynolds
stress have period $\Delta x_{\rm MRS} = 1/k_y\hat{s}$, only one such period is
shown. Note how the radial profile of the quasi-linear estimate of
$\partial^2\hat{\rm RS}_{k_y}/\partial x^2$ is narrow in the case of kinetic
electrons and localized around the MRS at $x=0$, clearly related to the fine
structures of the eigenmode at MRSs and the associated broad tail in ballooning representation. As expected, the corresponding radial
profile is much broader in the case of adiabatic electrons, as fine structures
at MRSs are essentially absent in this case.

%
%-------------------------------------------------------------------------
%    
\subsection{Evidence of zonal flow drive by modulational instability and 
self-interaction in reduced simulations}\label{Refreduced}
   
In this subsection, we consider reduced non-linear simulation setups in
tokamak geometry to clearly illustrate the two basic mechanisms driving zonal
flows discussed in \S\ref{Refslab} and \S\ref{Reftoroidal}. For these
reduced simulations, the same physical parameters as summarized in
Table~\ref{Parameterset} is considered, however with particular initial
conditions defined as follows: An unstable linear eigenmode (hereby called the
pump mode with $k_y=k_{y,{\rm pump}}\neq 0$) is initialized with an amplitude a few times ($\sim 5$) less than the corresponding one in
the fully saturated turbulence simulation discussed in \S\ref {Sec. SI in
  turbulent sims}. The purpose of the simulation is to study how this single
linear eigenmode drives zonal modes via either the modulational instability or
self-interaction. The eigenmode with $k_{x0}=0$ and $k_{y}\rho_i=0.28$ was
chosen for this pump mode, as it is among the most linearly unstable ones and
also contributes significantly to the non-linear fluctuation spectra
in the fully developed turbulence simulations (see
figures~\ref{gammavsky_kinadcomp} and \ref{Qivskymin}(b)). In addition, zonal
Fourier modes (with $k_y=0$) are initialized to amplitudes $10-12$ orders of
magnitude less than the pump mode to provide a necessary seed for possible
modulational instabilities. All Fourier modes not part of the pump and zonal
modes are initialized to zero. With this initial setup, only $k_y$ modes which
are harmonics of $k_{y,{\rm pump}}$ can possibly develop non-linearly ($k_y =
p\,k_{y,{\rm pump}}$, $p\in\mathbb{N}$). For these reduced simulations we
therefore set $k_{y, \min}=k_{y,{\rm pump}}$ and actually only considered
$N_{k_y}=8$ Fourier modes. The flux-tube width in the $x$ direction was set to
$L_x=M/\hat{s}k_{y, \min}$ with $M=32$, so that it remains the same as in full
turbulence simulations and the resulting fine $k_x$-spectrum allows for a
detailed analysis of $k_x-$dependence of zonal mode growth. The remaining
numerical resolutions are kept the same as in Table~\ref{Parameterset}. This
system is then let to evolve until higher $k_y$ Fourier harmonics start to
develop amplitudes similar to the fundamental $k_{y,{\rm pump}}$. These steps
ensure that the dominant non-linear interactions mainly involve only $k_y = 0$ and
$k_{y,{\rm pump}}$. This reduced non-linear setup thus clearly isolates the
contribution to the drive of zonal modes from a single $k_y$ mode, while in
the much more complex case of a standard fully developed turbulence
simulation, multiple $k_y$ contributions may act simultaneously.
    
\begin{figure}
\centering
\includegraphics[width=0.49\columnwidth]{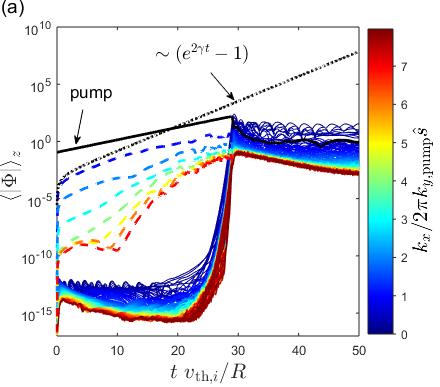}
\includegraphics[width=0.49\columnwidth]{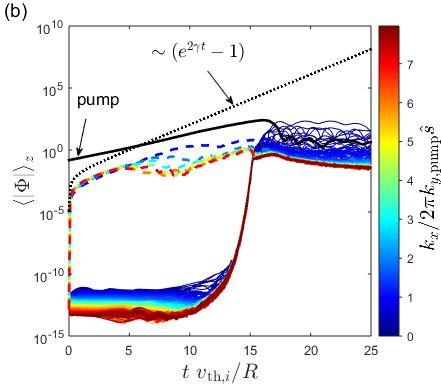}
\caption{Lin-log plots. Coloured lines represent the time evolution of the
  $z$-averaged amplitudes $\langle|\hat{\Phi}_{k_x,k_y=0}|\rangle_z$ of zonal
  modes in reduced non-linear simulations initialized with a single large
  amplitude $k_{y, {\rm pump}}\neq 0$ eigenmode, considering either (a)
  adiabatic or (b) kinetic electrons. The colourbar maps the value of $k_x$ in
  units of $2\upi k_{y, {\rm pump}}\hat{s}$. Modes with $k_x=p'\,2\upi k_{y,
    {\rm pump}}\hat{s}$, $p'\in\mathbb{N}$, are plotted with dashed lines. The
  solid black line represents the time trace of the pump Fourier mode
  amplitude $\langle|\hat{\Phi}_{k_x=0,k_{y,{\rm {\rm pump}}}}|\rangle_z$. The
  dotted black line represents an evolution proportional to $(e^{2\gamma
    t}-1)$, where $\gamma$ is the linear growth rate of the pump mode.}
\label{init_absPhivst}
\end{figure}

Figure~\ref{init_absPhivst} plots the evolution of the pump mode and the zonal modes driven by it. Solid black line represents the time trace of the $z$-averaged amplitude of the most dominant Fourier mode $\langle|\hat{\Phi}_{k_{x0},k_{y,{\rm pump}}}|\rangle_z$(t) composing the pump mode. Note that the other linearly coupled Fourier modes composing the pump mode are not shown. Coloured lines represent the zonal mode amplitude $\langle|\hat{\Phi}_{k_x,k_y=0}(t)|\rangle_z$ for each $k_x$. These results
are shown for reduced non-linear simulations considering either adiabatic electrons (figure~\ref{init_absPhivst}(a)) or kinetic
electrons (figure~\ref{init_absPhivst}(b)). 

In the initial stage, the pump eigenmode grows exponentially with
corresponding linear growth rates, {\it i.e.} $\gamma R/v_{{\rm th},i}=0.25$
in the case of adiabatic electrons and $\gamma R/v_{{\rm th},i} = 0.47$ in the
case of kinetic electrons.
    
\begin{figure}
\centering
\includegraphics[width=0.49\columnwidth]{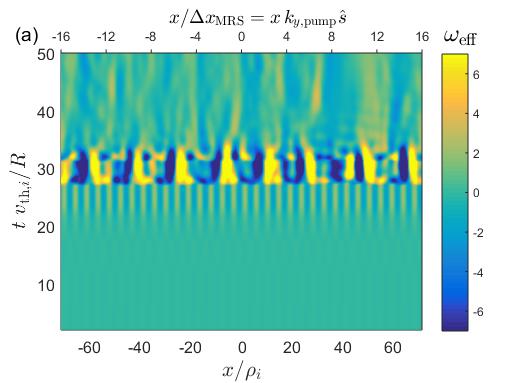}
\includegraphics[width=0.49\columnwidth]{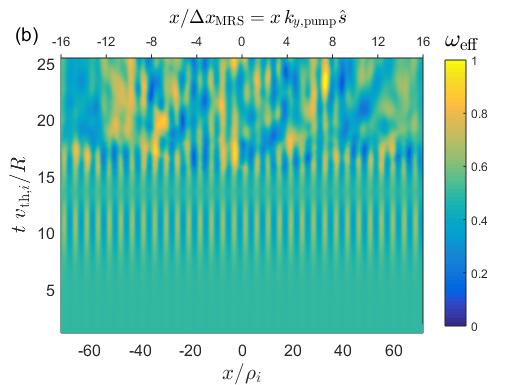}
\caption{Effective zonal flow shearing rate
  $\omega_{\rm eff}$ as a function of radial position $x$ and time $t$ for the same reduced non-linear simulations as in
  figure~\ref{init_absPhivst}. Cases
  with either (a) adiabatic or (b) kinetic electrons are shown.}
\label{init_wonxandt}
\end{figure}

Zonal Fourier modes $\hat{\Phi}_{k_x',k_y=0}$ with $k_x'=p'\,2\upi k_{y,{\rm
    pump}}\hat{s}$, $p'\in \mathbb{Z}$, (see dashed coloured time traces in
figure~\ref{init_absPhivst}) are driven by the large amplitude pump eigenmode
through the self-interaction mechanism, {\it i.e.} are driven by multiple
quadratic non-linearities, each involving two exponentially growing $k_x$
Fourier components of the pump , $\hat{\Phi}_{k_x,k_{y,{\rm pump}}}(t)$,
$\hat{\Phi}_{k_x'',k_{y,{\rm pump}}}(t) \sim e^{\gamma t}$, where
$k_x^{(\prime\prime)}=k_{x0}+p^{(\prime\prime)}2\upi k_{y,{\rm pump}}\hat{s}$,
$p^{(\prime\prime)}\in \mathbb{Z}$, and $p-p''=p'$. One thus has for
$k_x'=p'\,2\upi k_{y,{\rm pump}}\hat{s}$:
\begin{align*}
  \hat{\Phi}_{k_x',k_y=0}(t)
  -
  \hat{\Phi}_{k_x',k_y=0}(0)
  \,\sim\,&
  \int_0^t
  \hat{\Phi}_{k_x,k_{y,{\rm pump}}}(t')\,
  \hat{\Phi}^*_{k_x'',k_{y,{\rm pump}}}(t')\,
  dt'
  \,\sim\,
  \int_0^t e^{2\gamma t'}dt' \\
  \,\sim\,&
  (e^{2\gamma t}-1).
\end{align*}
Note that for $t \ll 1/\gamma$ these modes thus start by growing linearly in
time.  In the initial stage of the simulations shown in
figures~\ref{init_absPhivst}(a) and \ref{init_absPhivst}(b), \emph{i.e.} for $t
\lesssim 28 R/v_{{\rm th},i}$ and $t \lesssim 15 R/v_{{\rm th},i}$ in the case
of adiabatic and kinetic electrons respectively, the Fourier modes driven by
self-interaction, $\hat{\Phi}_{p'\,2\upi k_{y,{\rm pump}}\hat{s},k_y=0}$,
therefore dominate the zonal $k_x$-spectrum. Black dotted lines in these
figures indicate the time evolution $\sim (e^{2\gamma t}-1)$, 
providing a good fit to the initial evolution of these particular zonal
modes.  As discussed in \S\ref{Reftoroidal}, these modes are driven by the
Reynolds stress resulting from the self-interaction of the linear eigenmode
with $k_y = k_{y,{\rm pump}}$, which in direct space is periodic in $x$ with
period $\Delta x_{\rm MRS} = 1/k_{y,{\rm pump}} \hat{s}$ and aligned with
corresponding MRSs. This is clearly reflected in Figures~\ref{init_wonxandt}(a)
and \ref{init_wonxandt}(b), plotting the effective zonal flow shearing rate
$\omega_{\rm eff}$ as a function of $x$ and time $t$, again for simulations
with either adiabatic or kinetic electrons. At least in the initial stage of the
simulations, these plots indeed present a periodic radial variation of
$\omega_{\rm eff}$ with period $1/k_{y,{\rm pump}} \hat{s}$ and perfectly
aligned with the MRSs of the pump mode.

%% In the initial stage of the simulations shown in figure~\ref{init_absPhivst},
%% \emph{i.e.} for $t < 30 R/v_{{\rm th},i}$ and $t < 15 R/v_{{\rm th},i}$ in the
%% case of adiabatic and kinetic electrons respectively, Fourier modes driven by
%% self-interaction dominate the zonal $k_x$-spectrum. As discussed in
%% Sec. \ref{Reftoroidal}, the Reynolds stress drive of zonal modes from a given
%% $k_y$ linear eigenmode via self-interaction is periodic in $x$ with period
%% $\Delta x_{\rm MRS} = 1/k_y\hat{s}$ and aligned with corresponding MRSs. This is clearly
%% reflected in figure~\ref{init_wonxandt}, plotting the effective zonal flow
%% shearing rate $\omega_{\rm eff}$ as a function of $x$ and time $t$, again for
%% simulations with either adiabatic or kinetic electrons.
    
Zonal Fourier modes $\hat{\Phi}_{k_x',k_y=0}$ with $k_x'\neq p'\,2\upi
k_{y,{\rm pump}}\hat{s}$, $p'\in\mathbb{Z}$, (see solid coloured time traces in
figure~\ref{init_absPhivst}) are however driven by the pump eigenmode via the
modulational instability mechanism, \emph{i.e.} mainly through the quadratic
non-linearities involving the large amplitude Fourier component
$\hat{\Phi}_{k_{x0},k_{y,{\rm pump}}}$ of the pump and either one of the
initially low amplitude sideband modes $\hat{\Phi}_{k_{x0}\pm k_x',k_{y,{\rm
      pump}}}$. As a result, these particular zonal modes grow as $\sim
e^{\gamma_{\rm mod} t}$, where $\gamma_{\rm mod}$ stands for the growth rate
of the modulational instability. Given that $\gamma_{\rm mod} \sim|\Phi_{{\rm
    pump}}|$ \citep{Hasegawa1979} and that the pump amplitude itself grows
exponentially, $|\Phi_{{\rm pump}}| \sim e^{\gamma t}$, these zonal modes
effectively end up growing super-exponentially.
    
One can also see in figure~\ref{init_absPhivst}(b) that all plotted zonal modes
driven by self-interaction, $\hat{\Phi}_{p'2\upi k_{y,{\rm
      pump}}\hat{s},k_y=0}$, $p'=1, \ldots, 7$, are from the start driven up
to similarly large amplitudes in the case of kinetic electrons, as opposed to
the adiabatic case in figure~\ref{init_absPhivst}(a), where these same modes are
driven more weakly. This is explained by the different linear eigenmode
structures of the pump in the adiabatic and kinetic electron cases (see
corresponding ballooning representations in figure~\ref{Ballooning}).  With
kinetic electrons, the Fourier mode components $\hat{\Phi}_{k_x,k_{y,{\rm
      pump}}}(z)$, where $k_x=k_{x0}+ p\,2\upi k_{y,{\rm pump}}\hat{s}$ and
$p\in\mathbb{Z}^*$ (set of integers excluding 0), which compose the tail of the ballooning
representation, have much higher relative amplitudes compared to the main
Fourier component $\hat{\Phi}_{k_{x0},k_{y,{\rm pump}}}(z)$ than in the case
with adiabatic electrons. As a result, the self-interaction drive of a zonal
mode $\hat{\Phi}_{p'2\upi k_{y,{\rm pump}}\hat{s},k_y=0}$, $p' \in
\mathbb{Z}^*$, dominated by the non-linear interaction between
$\hat{\Phi}_{k_{x0},k_{y,{\rm pump}}}(z)$ and $\hat{\Phi}_{k_{x0} + p'2\upi
  k_{y,{\rm pump}}\hat{s},k_{y,{\rm pump}}}(z)$, is stronger with kinetic then
with adiabatic electrons.

The weaker drive of the zonal modes by the self-interaction
mechanism in the case with adiabatic electrons also explains why  these
particular zonal modes end up getting overwhelmed by the ones driven by the
modulational instability mechanism (reflected by the absence of fine
structures at MRSs on $\omega_{\rm eff}$ in figure~\ref{init_wonxandt}(a) for $t
\gtrsim 28\,R/v_{{\rm th},i}$), while they remain significant in the case with
kinetic electrons even after saturation of the modes driven by the
modulational instability (reflected by the persistent fine structures on
$\omega_{\rm eff}$ in figure~\ref{init_wonxandt}(b) even for $t \gtrsim
15\,R/v_{{\rm th},i}$).

In the following two subsections, evidence of the self-interaction and
modulational instability mechanisms are provided in the case of fully
developed turbulence simulations, and their relative importance are compared
between the cases of adiabatic and kinetic electron responses. Such
simulations involve a fully saturated spectra of multiple $k_y$ modes
than the case in the just considered reduced non-linear simulations. We find
that the results obtained in the reduced non-linear simulations follow in
general to the full turbulence scenarios, but with certain differences.

%
%-------------------------------------------------------------------------
%
\subsection{Evidence of self-interaction in turbulence simulations}
\label{Sec. SI in turbulent sims}
In this subsection, the self-interaction mechanism is analysed in
fully developed turbulence simulations with physical and numerical parameters given in Table~\ref{Parameterset}.
We first test if in this non-linear system, the relative phase differences between the linearly
coupled $k_x-$Fourier modes remain the same as in corresponding linear eigenmode, as was assumed earlier in \S\ref{Reftoroidal}.
It is found that the relative phase difference is only partially preserved, and the consequent effect on the Reynolds stress 
driving zonal flows via self-interaction is illustrated.

\begin{figure}
\centering
  \includegraphics[width=0.44\columnwidth] {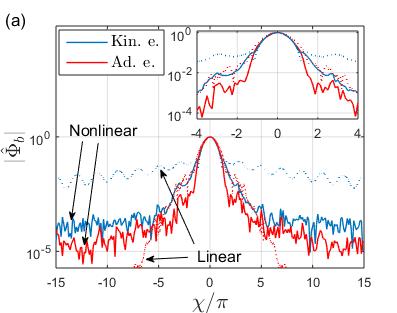}
  \includegraphics[width=0.55\columnwidth] {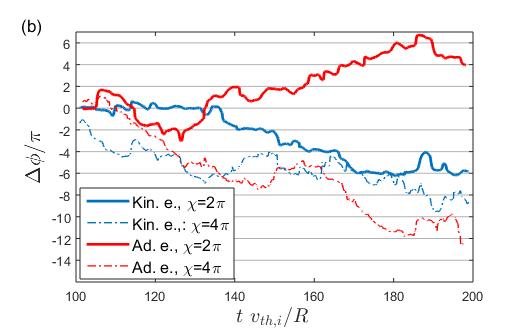}  
  \caption{(a) 
  Solid lines denote the absolute value of the time-averaged ballooning
structure of the electrostatic potential
$|\langle\hat{\Phi}_{b,{\rm nl}}(\chi,t)/\hat{\Phi}_{b,{\rm nl}}(\chi_0=0,t)\rangle_{t}|$
for $k_{x0}=0$ and $k_y\rho_i=0.28$ in the non-linear simulation described in
table~\ref{Parameterset}, with $k_{y,\min}\rho_i=0.035$ and $R/L_{T,i}=6$. Dotted lines denote the corresponding ballooning representation in linear simulations (same as in figure~\ref{Ballooning}). Inset figure shows the zoom near $\chi=0$. Red and blue colours represent adiabatic and kinetic electron simulations respectively. (b) Phase difference $\Delta\phi(\chi,t)=(\phi[(\hat{\Phi}_{b,{\rm nl}}(\chi,t)/\hat{\Phi}_{b,{\rm nl}}(\chi_0=0,t))(\hat{\Phi}_{b,{\rm lin}}(\chi_0=0)/\hat{\Phi}_{b,{\rm lin}}(\chi))])$ plotted as a function of time. Here, $\hat{\Phi}_{b,{\rm nl}}$ and $\hat{\Phi}_{b,{\rm lin}}$ denote the ballooning representation of the electrostatic potential for the same eigenmode considered in subplot (a), in non-linear and linear simulations respectively. Solid and dotted lines represent $\chi=2\upi$ and $4\upi$ respectively.}  
  \label{BalPhasediff}
\end{figure}

To find the non-linear modification of an eigenmode, first the absolute value of the time-averaged ballooning
structure of the electrostatic potential
$|\langle\hat{\Phi}_{b,{\rm nl}}(\chi,t)/\hat{\Phi}_{b,{\rm nl}}(\chi_0=0,t)\rangle_{t}|$
for an eigenmode with $k_{x0}=0$ and $k_y\rho_i=0.28$ in the full turbulence simulation is plotted along with the corresponding linear
ballooning structure
$|\hat{\Phi}_{b,{\rm lin}}(\chi)/\hat{\Phi}_{b,{\rm lin}}(\chi_0=0)|$, in figure~\ref{BalPhasediff}(a). See equation~(\ref{ballooning defs.}) for the definition of the ballooning structure
$\hat{\Phi}_{b}(\chi)$. The non-linear result with kinetic electrons starts
significantly deviating from the linear result for $|\chi|\gtrsim \upi$. That
is, the 'giant tails' in the ballooning structure are reduced compared to the
linear case. Nonetheless, the non-linear ballooning structure with kinetic
electrons retain more significant tails compared to the case with adiabatic
electrons. One also finds that, with kinetic electrons, the
relative phase along the ballooning structure
$\delta\phi_{\rm nl}(\chi,t)=\phi[\hat{\Phi}_{b,{\rm nl}}(\chi,t)/\hat{\Phi}_{b,{\rm nl}}(\chi_0=0,t)]$
remains approximately constant in time (contrary to adiabatic electron
results) and equal to its linear value
$\delta\phi_{\rm lin}(\chi)=\phi[\hat{\Phi}_{b,{\rm lin}}(\chi)/\hat{\Phi}_{b,{\rm lin}}(\chi_0=0)]$;
Here, $\phi[A]$=arg($A)$ stands for the phase or argument of the complex number $A$. This is true in particular for
$|\chi|\lesssim 4\upi$, as demonstrated by the solid blue line in
figure~\ref{BalPhasediff}(b) representing
$\Delta\phi(\chi,t)=\delta\phi_{\rm nl}(\chi,t)-\delta\phi_{\rm lin}(\chi)$ for
$\chi=2\upi$ in the kinetic electron case. While there are jumps of $2\upi$, the
line closely adheres to the linear phase difference. To be quantitative, one considers the estimate ${\rm MOD(\Delta\phi(\chi,t))}$,
where $ {\rm MOD}(A)=(\langle|{\rm mod}_{2\upi}(A)|^2\rangle_t)^{1/2}$, mod$_{2\upi}(A) \equiv A-2\upi \times {\rm round}(A/2\upi)$,
$A\in\mathbb{R}$, and the function round provides the nearest integer. One notes that, lower 
the value of MOD$(\Delta\phi)$, more strongly is the relative phase fixed by the linear coupling. Further, for uniform random values of $\Delta\phi$ between $-\upi$ and $\upi$, one gets MOD$(\Delta\phi)=0.58\upi$.  
It is found that, for the case with kinetic electrons,  ${\rm MOD}(\Delta\phi(\chi=2\upi,t))=0.34\upi$.
At $\chi=4\upi$, as illustrated by
the dotted blue line in figure~\ref{BalPhasediff}(b), the phase imposed by
linear coupling appears to be weakened but still present, giving MOD$(\Delta\phi(\chi,t))=0.50\upi$. The corresponding results with adiabatic electrons, denoted by the solid and
dashed red lines for $\chi=2\upi$ and $4\upi$ respectively, clearly do not retain
constant phase differences. The respective values of MOD$(\Delta\phi(\chi,t))$ are $0.57\upi$ and $0.61\upi$. 
    
In other words, in a full turbulence simulation with adiabatic electrons, for
a linearly unstable eigenmode with given $k_{x0}$ and $k_y$, linear coupling between the Fourier modes
$(k_x=k_{x0}+2\upi p k_y\hat{s},k_y)$ where $p\in\mathbb{Z}$
have become subdominant compared to non-linear coupling effects. Thus, each $(k_x,k_y)$ Fourier mode can be considered linearly decoupled as in a shearless slab system and hence, modulational instability is the only driving mechanism of zonal flows in
turbulent simulations with adiabatic electrons. On the other hand, with
kinetic electrons, for a linearly unstable physical mode with given $k_{x0}$ and
$k_y$, linear coupling of Fourier modes is preserved to some extent but reduced to only the three Fourier modes with $k_x=k_{x0}+2\upi pk_y\hat{s}$ and
$|p|=0,\pm 1$. Only these Fourier modes have their relative phases largely fixed by their linear dynamics
(and not for all $p\in\mathbb{Z}$, as assumed in \S\ref{Reftoroidal}).  Hence the self-interaction mechanism is
effectively reduced to driving the zonal Fourier modes with
$k'_x=\pm 2\upi k_y\hat{s}$. Furthermore, the self-interacting contribution to Reynolds stress
$\widetilde{\rm RS}_{k_{x0},k_y}(x)$ (defined in equation~(\ref{EqRSqlin})) from
the particular physical eigenmode, becomes essentially sinusoidal in x, spanning the full distance between corresponding MRSs with a period
of $\Delta x_{\rm MRS}=1/k_y\hat{s}$, and with fixed phase/sign in time. Note that the radial width of the fine-structures do not have a dependence on electron-ion mass ratio in turbulence simulations as they are already broadened to span the distance between MRSs as a result of non-linear coupling effects. This is unlike in linear simulations where where such a dependence exists \citep{Dominski2015}.

The net contribution RS$^{\rm si}$ from the
self-interaction mechanism to Reynolds stress can be obtained by summing $\widetilde{\rm
  RS}_{k_{x0},k_y}(x)$ over all $k_y$'s and all ballooning angles (measured by
$k_{x0}$; see (\ref{kx0deff})):
\begin{equation}
  {\rm RS}^{\rm si}(x)=\sum_{k_y>0}\ \hat{\rm RS}^{\rm si}_{k_y}(x)
\end{equation}
where
\begin{equation}
  \hat{\rm RS}^{\rm si}_{k_y}=\sum_{k_{x0}=-\upi k_y\hat{s}}^{\upi k_y\hat{s}}\widetilde{{\rm RS}}_{k_{x0},k_y}(x).    
\end{equation}
In figure~\ref{Lind2RS}, the self-interaction contribution $\partial^2{\rm
  RS}^{\rm si}/\partial x^2$ from a particular $k_y$ in full turbulence simulation is plotted. The
dashed lines denote the time average of $\partial^2\hat{\rm RS}^{\rm si}_{k_y}/\partial x^2$ for $k_y\rho_i=0.28$, normalised by the RMS in time of total $\partial^2\hat{\rm RS}_{k_y}/\partial x^2$, in simulations with
kinetic (blue) and adiabatic (red) electrons; the total Reynolds stress contribution ${\rm \hat{RS}}_{k_y}$ from a given $k_y$ being defined in (\ref{RS_ky in kx-repr}). This diagnostic simultaneously
measures the relative importance of self-interaction drive of zonal flows with respect to the total contribution to
Reynolds stress drive from the considered $k_y$, as well as how good its phase/sign is
fixed in time. One notes that, since for a given $k_y$, eigenmode with $k_{x0}=0$ is the most unstable/dominant, similar results
are obtained for $\widetilde{\rm RS}_{k_{x0}=0,k_y}(x)$ as well, instead of $\hat{\rm RS}^{\rm si}_{k_y}$ in this diagnostic.
From figure~\ref{Lind2RS}, one can conclude that, in full turbulence simulations with kinetic electrons, each $k_y$ contribution $\partial^2\hat{\rm RS}^{\rm si}_{k_y}/\partial x^2$
from self-interaction is significant. Furthermore, it is a
sinusoidal with a period of $1/k_y\hat{s}$ and maintains the same sign at each
radial position (although with randomly varying amplitude in time, as will be
shown at the end of \S\ref{Sec. MI in turbulent sims}). In particular, $\partial^2\hat{\rm RS}^{\rm si}_{k_y}/\partial x^2$ is zero at MRSs, negative to the left and positive to the right.
Whereas with adiabatic electrons, the
self-interaction contribution is weak, does not have a fixed sign, and essentially
averages out to zero over time. Further, with kinetic electrons, these
significant time-averaged self-interaction contributions, localized at MRSs of
each $k_y$, align at LMRSs, driving time-averaged zonal flows at these radial
positions. Whereas,
away from LMRSs, the time-averaged self-interaction contributions
$\partial^2\hat{\rm RS}^{\rm si}_{k_y}/\partial x^2$ from each $k_y$ are spatially
misaligned and cancel each other out, resulting in relatively negligible time
averaged zonal flow levels.
    
\begin{figure}
\centering
\includegraphics[width=0.6\columnwidth] {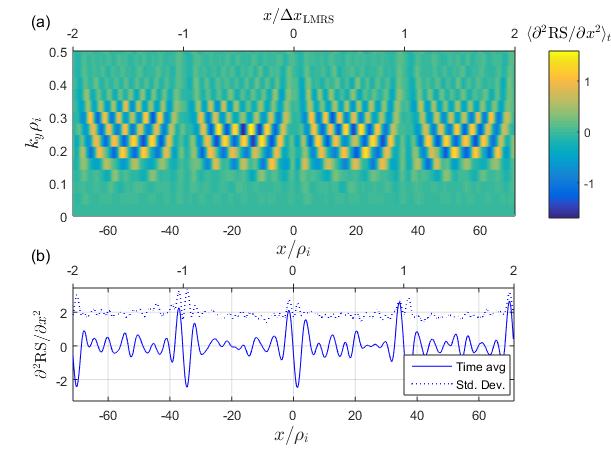}
  \caption{(a) Each $k_y$ contribution of $\langle\partial^2{\rm RS}/\partial x^2\rangle_{t}$ plotted as a function of the radial coordinate $x$. (b) Total $\langle\partial^2{\rm RS}/\partial x^2\rangle_{t}$ plotted vs $x$. Dashed line denote the standard deviation. Results correspond to the non-linear simulation described in
table~\ref{Parameterset}, with $k_{y,\min}\rho_i=0.035$, $R/L_{T,i}=6$ and kinetic electrons.}
  \label{d2RS}
\end{figure}

    In figure~\ref{d2RS}, the time average of the total $\partial^2\hat{\rm RS}_{k_y}/\partial x^2$ is plotted 
	as a function of both position $x$ and wavenumber $k_y$. While between LMRSs, $\partial^2\hat{\rm RS}_{k_y}/\partial x^2$ follows 
	essentially the same spatial orientation (phase) as 
	$\partial^2\hat{\rm RS}^{\rm si}_{k_y}/\partial x^2$ in figure~\ref{Lind2RS}, near LMRSs, 
	there is a reversal of spatial phase. We suspect this to be a result of the back reaction 
	of zonal modes driven by self-interaction on turbulence. At the moment, we postpone further 
	analysis on this to a later time.
	
\begin{figure}
\centering
\includegraphics[scale=0.38]{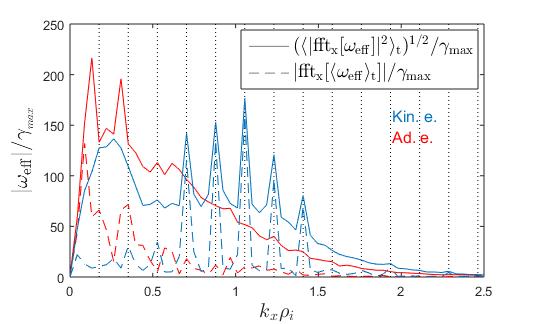}
\caption{$k_x$ spectra of the effective shearing rate $\omega_{\rm eff}$ in turbulence simulations with adiabatic (red) and kinetic (blue) electrons. Corresponding parameters are given in table~\ref{Parameterset}, with $k_{y, \min}\rho_i=0.035$ and $R/L_{T,i}=6$. Solid lines and dashed lines represent time average of the amplitude and amplitude of the time average of $\omega_{\rm eff}$ respectively. All traces are normalised by corresponding maximum linear growth rate $\gamma_{\max}$. Vertical dotted lines indicate positions where $k_x=2\upi p\hat{s}k_{y, \min}$; $p\in\mathbb{Z}$.}
\label{w_kxspectra}
\end{figure}

    Recall from earlier in this subsection that in kinetic electron turbulence simulations, the self-interaction contribution from a given $k_y\neq 0$ is
effectively reduced to driving the zonal Fourier modes with
$k'_x=\pm 2\upi k_y\hat{s}$. The combined self-interacting 
	contributions from the various $k_y$'s therefore drive zonal Fourier modes with 
	$k_x=2\upi p\hat{s}k_{y, \min}$, where $p\in\mathbb{Z}$. This can be seen as peaks in the 
	$k_x$-spectra of the time-averaged shearing rate $\langle\omega_{\rm eff}\rangle_{t}$, 
	shown with blue dashed line in figure~\ref{w_kxspectra} and correspond to the time-constant 
	structures at LMRSs separated by $\Delta x_{LMRS}=1/\hat{s}k_{y, \min}$ in 
	figure~\ref{omegaecrossbionsvsx_sepspcompare}. Given that, in the case of the reference turbulence simulation with kinetic electrons, the highest contribution to heat flux 
	(see figure~\ref{Qivskymin}(b)) and the $|\Phi|^2$ amplitude spectra is at $k_y\rho_i\sim0.21$ and 
	that for a given $k_y$, eigenmodes with $k_{x0}=0$ are the most dominant,
	the zonal mode with $k_x\rho_i\sim 2\upi\hat{s}\times 0.21\sim 1.06$ are driven strongly via the self-interaction mechanism and hence
	has the highest value in the $k_x$-spectra of shearing rate. Similarly, 
	those $k_y$'s with lower heat flux and amplitude contributions, contribute lesser towards 
	the self-interaction mechanism and thereby explains the number and position of peaks in 
	the $k_x$-spectra of shearing rate with kinetic electrons in figure~\ref{w_kxspectra}.
	
    Solid red line in figure~\ref{w_kxspectra} denotes the $k_x$-spectra of $\omega_{\rm eff}$ in the 
	simulation with adiabatic electron response. These zonal Fourier modes are driven predominantly by modulational instability. In the next subsection, we test, if consequently, as a result of the frequency matching involved in the modulational instability mechanism, 
	those zonal modes having high shearing rate amplitudes end up playing a major role in pacing 
	the dynamics of much of the $(k_x,k_y\neq 0)$ Fourier modes in adiabatic electron simulations. 
	\\
   
%
%------------------------------
% 
\subsection{Evidence of modulational instability in turbulence simulations. Bicoherence and correlation analysis.}
\label{Sec. MI in turbulent sims}

In this subsection, we verify in fully developed turbulence simulations, that indeed with adiabatic electron response, zonal flows are driven predominantly via modulational instability, as found in reduced simulations in \S\ref{Refreduced} and as discussed in \S\ref{Sec. SI in turbulent sims}. We test this by doing a bicoherence-like analysis. We simultaneously show that, as a consequence, this leads to \emph{correlated} contributions from the various $k_y$'s to the Reynolds stress drive of zonal modes. We furthermore show that in kinetic electron simulations, the self-interaction contributions to Reynolds stress from the various $k_y$'s are random in time and \emph{decorrelated} with each other.

As already discussed earlier, modulational instability mechanism involves resonant 3-wave interactions which require frequency matching. The strength of a particular resonant interaction between the 3 Fourier modes, ${\bf k}=(k_x,k_y)$, the zonal mode ${\bf k'}=(k'_{x},0)$ and daughter mode ${\bf
  k''}={\bf k}-{\bf k'}=(k_x-k'_x,k_y)$, can thus be measured via a bicoherence-type analysis, essentially involving the time-average of the triplet product
    \begin{equation}
        T({\bf k}\ ;+{\bf k'})=\hat{\Phi}_{\bf k}(t)\hat{\Phi}^*_{\bf k'}(t)\hat{\Phi}^*_{\bf k''}(t)
    \end{equation}
where  $\hat{\Phi}_{\bf q} (t) \sim {\rm exp}[-i(\omega_{\bf q} t + \phi_{\bf q})]$ is the complex time dependent amplitude of the electrostatic field with any Fourier mode {\bf q}, having a frequency $\omega_{\bf q}$, initial phase $\phi_{\bf q}$, and evaluated at z=0. If the Fourier modes $[\mathbf k, \mathbf k', \mathbf k'' = \mathbf k-\mathbf k']$ are frequency matched in the simulation time, i.e. $\omega_{\bf k} = \omega_{\bf k'} + \omega_{\bf k''}$, then $\langle T({\bf k}\ ;+{\bf k'})\rangle_t\neq 0$, where $\langle .\rangle_t$ stands for the time-average over the simulation time.

Here, we remark that, for a given set of Fourier modes $[\mathbf k, \mathbf k', \mathbf k'' = \mathbf k-\mathbf k']$ under frequency matching condition, different resonant decay mechanisms could be possible, each with a specific phase difference $\Delta\phi=\phi_{\bf k}-\phi_{\bf k'}-\phi_{\bf k''}$ associated to it. Hence, a non-zero time average $\langle T({\bf k}\ ;+{\bf k'})\rangle_t$ over the simulation time is indicative of a particular resonant decay mechanism being persistent throughout the simulation.

Now, a normalised measure of the strength of the resonant 3-wave interaction can be calculated by the following estimate: 
    \begin{equation}
        b_{N}({\bf k}\ ;+{\bf k'}) = \frac{|\langle T({\bf k}\ ;+{\bf k'})\rangle_t|}{\langle |T({\bf k}\ ;+{\bf k'})|\rangle_t},
    \end{equation}
defined as the bicoherence between the Fourier triplet $[\mathbf k, \mathbf k', \mathbf k'' = \mathbf k-\mathbf k']$. Note $0\leq b_{N}\leq 1$. Further, $b_{N}({\bf k}\ ;+{\bf k'})\simeq 1$ indicates a fully resonant 3-wave interaction between ${\bf k}$, ${\bf k''}$ and the zonal mode ${\bf k'}$, while $b_{N}\simeq 0$ indicates a non-resonant process. Since modulational instability is a simultaneous resonant interaction between both triplets $[\mathbf k, \mathbf k', \mathbf k'' = \mathbf k-\mathbf k']$ and $[\mathbf k, -\mathbf k', \mathbf k''' = \mathbf k+\mathbf k']$, we define total bicoherence
    \begin{equation}
        B_{N}({\bf k}\ ;{\bf k'}) = (b_N({\bf k}\ ;+{\bf k'})+b_N({\bf k}\ ;-{\bf k'}))/2,
        \label{Eq_bicoh}
    \end{equation} 
such that a value of $B_{N}\simeq 1$ indicates a fully resonant modulational instability mechanism. In general, values of $B_N$ closer to zero are indicative of non-resonant interactions while larger values of $B_{N}$ are characteristic of modulational instability mechanisms.

\begin{figure}
\centering
\includegraphics[scale=0.44]{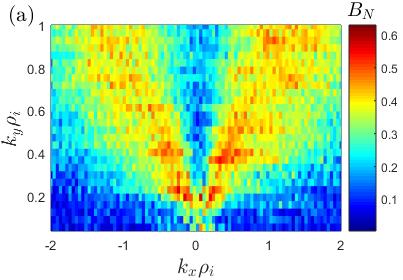}
\hspace{2mm}
\includegraphics[scale=0.44]{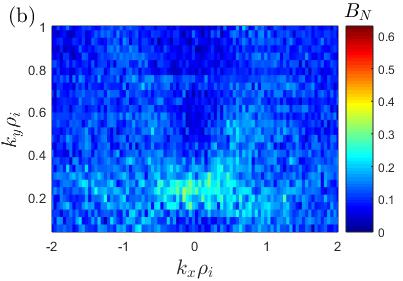}
\caption{The bicoherence level $B_N$, as defined by equation~(\ref{Eq_bicoh}), in (a) adiabatic and (b) kinetic electron turbulence simulations, for the zonal modes with $k'_{x}\rho_i=0.13$ and $0.26$ respectively. Results correspond to simulations with parameters as given in table~\ref{Parameterset}, with $k_{y, \min}\rho_i=0.035$ and $R/LT_i=6$.}
\label{Energy_bicoh}
\end{figure}

Figure~\ref{Energy_bicoh}(a) shows $B_N({\bf k}\ ;{\bf k'})$ as a function of ${\bf k}=(k_x,k_y)$, in the reference simulation with adiabatic electrons, for the fixed zonal mode with ${\bf k'}=(k'_{x}=0.13\rho_i^{-1},0)$ which has the highest contribution to the zonal shearing rate $k_x$-spectra (see solid red line in figure~\ref{w_kxspectra}). One can see that most ${\bf k}=(k_x,k_y)$ have high values of $B_N\gtrsim 0.3$, reflecting the dominance of resonant 3-wave interaction processes. Whereas in figure~\ref{Energy_bicoh}(b), corresponding kinetic electron simulation result for the zonal mode with ${\bf k'}=(k'_{x}=0.26\rho_i^{-1},0)$ having highest contribution to the shearing rate $k_x$-spectra shows much smaller values of $B_N$, indicating much weaker resonant 3-wave interactions. Similar differences are seen between adiabatic and kinetic electron results for other zonal modes ${\bf k'}=(k'_{x},0)$ as well.

\begin{figure}
\centering
\includegraphics[scale=0.45]{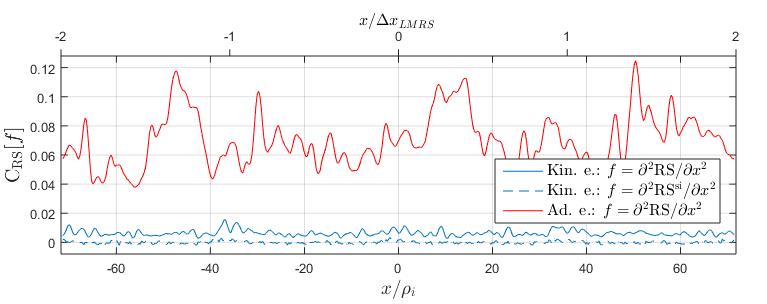}
\caption{Correlation between the $k_y$ modes of $\partial^2{\rm RS}/\partial x^2$, as defined by equation~(\ref{Eq_normcorr}), as a function of $x$ in turbulence simulations with kinetic (solid blue line) and adiabatic (solid red line) electrons. Corresponding parameters are given in table~\ref{Parameterset}, with $k_{y, \min}\rho_i=0.035$ and $R/L_{T,i}=6$. Correlation between the $k_y$ modes of self-interacting contribution $\partial^2{\rm RS}^{\rm si}/\partial x^2$ in the simulation with kinetic electrons is shown with dashed blue line.}
\label{normcorr}
\end{figure}

Note that, for a given zonal mode ${\bf k'}=(k_x',0)$, under wave vector matching condition, one has the following approximate estimate between associated triple products and the Reynolds stress contributions driving the particular zonal mode:
\begin{align}
\label{TtoRSrelation}
\sum_{k_x}T({\bf k}\ ;{\bf k'}) \sim \hat{\Phi}_{\bf k'}^*\hat{\rm RS}_{k_y},
\end{align}
where $\hat{\rm RS}_{k_y}$ has been defined in equation~(\ref{RS_ky in kx-repr}). Now, if $\langle T({\bf k}\ ;{\bf k'})\rangle_t$ (and the corresponding total bicoherence $B_N({\bf k}\ ;{\bf k'})$) for multiple Fourier modes ${\bf k}=(k_x,k_y)$ are simultaneously  significant for the same dominant zonal mode ${\bf k'}$ (as is the case in figure~\ref{Energy_bicoh}(a) for adiabatic electron simulation), then based on (\ref{TtoRSrelation}) one can conclude that the Reynolds stress contributions $\hat{\rm RS}_{k_y}$ from the various $k_y$'s tend to be in phase with the particular zonal mode. This implies a coherent and thus correlated contributions (drives) from these various $\hat{\rm RS}_{k_y}$ ($\partial^2\hat{\rm RS}_{k_{y}}/\partial x^2$). To verify this explicitly, we define an effective correlation function ${\rm C_{RS}}$ measuring the average correlation between all pairs of [$\partial^2\hat{\rm RS}_{k_{y,1}}/\partial x^2$, $\partial^2\hat{\rm RS}_{k_{y,2}}/\partial x^2$] for $k_{y,1}\neq k_{y,2}$:
    \begin{equation}    
    {\rm C_{RS}}[f]= \sum_{\substack{k_{y,i},\ k_{y,j} \\ k_{y,j}>k_{y,i}}
    }\frac{{\rm Cov}[\hat{f}_{k_{y,i}},\hat{f}_{k_{y,j}}]}{\sigma[\hat{f}_{k_{y,i}}]\sigma[\hat{f}_{k_{y,j}}]} \ \Big/
    \sum_{\substack{k_{y,i},\ k_{y,j} \\ k_{y,j}>k_{y,i}}} 1\ .
    \label{Eq_normcorr}
    \end{equation}
$f=\partial^2{\rm RS}/\partial x^2$, $\hat{f}_{k_{y}}$= $\partial^2\hat{\rm RS}_{k_y}(x)/\partial x^2$, covariance Cov$[a,b]=(\sigma^2[a+b]-\sigma^2[a]-\sigma^2[b])/2$, and variance $\sigma^2[a]=\langle|a-\langle a\rangle_t|^2\rangle_t$, with $\langle .\rangle_t$ representing average over simulation time.  Note that ${\rm C_{RS}}\in [0,1]$, with 1 corresponding to perfect correlation between Reynolds stress drive from all $k_y$'s and 0 corresponding to total decorrelation between them.
In figure~\ref{normcorr}, C$_{\rm RS}$ is plotted as a function of the radial position x, for adiabatic and kinetic electron turbulence simulations. We can find that ${\rm C_{RS}}[\partial^2 {\rm RS}/\partial x^2]$ for adiabatic electron case (solid red line) only shows $\sim 7\%$ correlation. However the important point to note is that this is roughly an order of magnitude higher than that for kinetic electron simulation (solid blue line). Thus, this diagnostic verifies that in adiabatic electron turbulence simulations where modulational instability mechanism is stronger, Reynolds stress drive $\partial^2\hat{\rm RS}_{k_{y}}/\partial x^2$ from the various $k_y$'s are indeed more correlated with each other, when compared to kinetic electron simulations where modulational instability mechanism is weaker.

The same correlation diagnostic is now performed on the
self-interacting part of Reynolds stress in kinetic electron
simulations. In figure~\ref{normcorr}, the correlation between the
$k_y$ contributions to $\partial^2{\rm RS}^{\rm si}/\partial x^2$,
measured by ${\rm C_{RS}}[\partial^2 {\rm RS^{si}}/\partial x^2]$, is
found to be nearly zero. This implies that, each $k_y$ contribution to
$\partial^2{\rm RS}^{\rm si}/\partial x^2$ at a given radial position
(while in general having a fixed sign) has an amplitude that is
completely independent and decorrelated with each other. The
self-interaction drive of zonal modes from each $k_y$ thus acts like
random kicks. Consequently, if the decorrelated (\emph{incoherent})
contributions from the self-interaction mechanism becomes significant,
it may disrupt the (\emph{coherent}) modulational instability
interactions. We interpret this as being the cause for lower
bicoherence values of $B_N$ in figure~\ref{Energy_bicoh}(b), as well
as the lower level of $\rm C_{RS}[\partial^2 {\rm RS}/\partial x^2]$
in figure~\ref{normcorr} with kinetic electrons as compared to
adiabatic electrons.\\[0.2cm]

  To summarise, in this section~\ref{AnalyzingZFdrive},
  the two different mechanisms driving zonal flows, namely
  modulational instability and self-interaction, have been discussed
  in detail. We have shown that, as a result of the stronger linear
  $k_x$-couplings within an eigenmode in kinetic electron simulations,
  the associated drive of zonal flows via self-interaction is much
  more significant, than in the case of adiabatic electron
  simulations. This has been illustrated in both reduced and full
  turbulence simulations. Furthermore, in turbulence simulations, it
  was shown that the contributions to zonal flow drive from each $k_y$
  mode, via the self-interaction mechanism, are random and decorrelated
  with each other in time. In the next section, we explain how such a
  decorrelated drive can lead to a decrease in the fluctuating zonal
  flow levels with decreasing $k_{y,\min}\rho_i$.

%
%----------------------------------------------------------------------
%
\section{Estimating the reduction in zonal flow drive from self-interaction  with decreasing $k_{y, \min}\rho_i$}\label{Statdep}

The low temporal correlation between the various $k_y$ contributions to the Reynolds stress drive $\partial^2{\rm RS}/\partial x^2$ of zonal flows in kinetic electron simulations, which is a result of strong self-interaction mechanism, could explain the decrease in the level of shearing rate $\omega_{\rm eff}$ associated to fluctuating $E\times B$ zonal flows with increasing toroidal system size, over the range of  $k_{y,\min}\rho_i$ considered in figure~\ref{omegaeffvskymin}(c). This explanation is based on simple statistical arguments described as follows.    
    
To start with, let us assume that the fluctuation energy density, proportional to amplitude density $|\Phi|^2$ of the fluctuating electrostatic field $\Phi$, remains the same across the simulations with different $k_{y, \min}$s. Consequently the amplitude $|\hat{\Phi}_{k_y}|^2$ of each $k_y$ Fourier component of the electrostatic field scales as $1/N_y$ where $N_y$ is the number of participating modes, according to the following estimate based on Parseval's identity:
    \begin{equation}
    \label{phiscaling}
    \frac{1}{L_y}\int_0^{L_y}|\Phi|^2dy=\sum_{k_y=0}^{N_y}|\hat{\Phi}_{k_y}|^2=\rm{constant}
    \Longrightarrow|\hat{\Phi}_{k_y}|^2\sim\frac{1}{N_y}.
    \end{equation}
Here, $\hat{\Phi}_{k_y}$ is defined as per the relation
    \begin{equation}
        \Phi(x,y,z)=\sum_{k_y}\hat{\Phi}_{k_y}(x,z)e^{ik_yy}.
    \end{equation}
    
Since Reynolds stress is a quadratic quantity in $\Phi$, the scaling in relation~\ref{phiscaling} also implies that each $k_y$ contribution to the Reynolds stress drive scales as $\partial^2\hat{\rm RS}_{k_y}/\partial x^2\sim|\hat{\Phi}_{k_y}|^2\sim 1/N_y$. Since $N_y=k_{y,\max}/k_{y,\min}$ and $k_{y,\rm max}$ remains the same across simulations in the $k_{y,\rm min}$ scan, one also has $1/N_y\sim k_{y,\rm min}$. 

Now, further assuming the various $k_y$ contributions to Reynolds stress drive $\partial^2\hat{\rm RS}_{k_y}/\partial x^2$ to be nearly fully decorrelated (characteristic of kinetic electron simulations; see figure~\ref{normcorr}), the variance of the total Reynolds stress drive $\partial^2{\rm RS}/\partial x^2=\sum_{k_y}\partial^2\hat{\rm RS}_{k_y}/\partial x^2$ becomes the sum of variances of all $k_y$ contributions:
    \begin{equation}
    {\rm Var}\left(\sum_{k_y}\frac{\partial^2}{\partial x^2}\hat{\rm RS}_{k_y}\right)\simeq\sum_{k_y}{\rm Var}\left(\frac{\partial^2}{\partial x^2}\hat{\rm RS}_{k_y}\right)\\
    \sim N_y\frac{1}{{{N_y}^2}}\sim\frac{1}{N_y}\\
    \sim k_{y, \min},
    \label{RSscaling}
    \end{equation}
having made use of ${\rm Var}(\partial^2{\rm \hat{RS}}_{k_y}/\partial x^2)\sim 1/N_y^2$, given that $\partial^2{\rm \hat{RS}}_{k_y}/\partial x^2\sim 1/N_y^2$. We therefore expect the standard deviation ${\rm SD}$ of the Reynolds stress drive $\partial^2{\rm RS}/\partial x^2$ to scale as ${\rm SD}={\rm Var}^{0.5}\sim k_{y, \min}^{0.5}$. Further, since $\partial^2{\rm RS}/\partial x^2$ is a proxy for the drive of zonal flows, this explains a decrease in the standard deviation of $\omega_{\rm eff}$ with decreasing $k_{y, \min}$.     
    
\begin{figure}
\centering
  \includegraphics[width=0.45\linewidth]{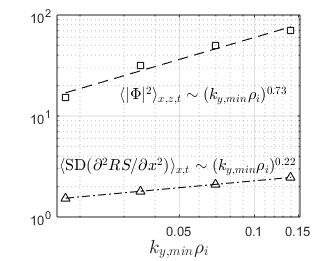}
  \caption{Triangles: Standard deviation $\langle{\rm SD}(\partial^2{\rm RS}/\partial x^2)\rangle_{x,t}$ of Reynolds stress drive plotted as a function of $k_{y, \min}$ in log-log scale, in turbulence simulations with kinetic electrons. Corresponding parameters are given in table~\ref{Parameterset}, with $R/L_{T,i}=6$. Fit (dashed-dotted line) shows scaling $\sim (k_{y,\min}\rho_i)^{0.22}$. Squares: $\langle|\hat{\Phi}|^2\rangle_{x,z,t}(k_y\rho_i=0.28)$ plotted as a function of $k_{y, \min}$, in the same set of simulations. Fit (dashed line) shows scaling $\sim(k_{y,\min}\rho_i)^{0.73}$}.
  \label{d2Restddevvskymin}
\end{figure}

From the $k_{y,\min}$ scan performed with kinetic electrons, $R/L_{T,i}=6$ and other parameters as given in Table~\ref{Parameterset}, we find that the x and time-averaged standard deviation $\langle{\rm SD}(\partial^2{\rm RS}/\partial x^2)\rangle_{x,t}$ of the Reynolds stress drive actually scales as $k_{y, \min}^{0.22}$, as shown in figure~\ref{d2Restddevvskymin}. This mismatch between the expected and observed scalings pushes us to reconsider the assumptions made that led to the estimate ${\rm SD}\sim k_{y, \min}^{0.5}$.
    
The small but non-zero value of the correlation between the $k_y$ contributions of $\partial^2{\rm RS}/\partial x^2$ in these simulations (see solid blue line in figure~\ref{normcorr}) may account for a small part of the discrepancy. However, the primary cause seems to arise from the faulty assumption of constant fluctuation energy density across simulations with different $k_{y,\rm min}$s. As the zonal flow shearing rate decreases with decreasing $k_{y, \min}$, the energy density in the system tends to increase, so that the estimate made in relation~\ref{phiscaling} has to be reconsidered. This is confirmed in figure~\ref{d2Restddevvskymin} where the same set of simulations show $\langle|\hat{\Phi}_{k_y}|^2\rangle_{x,z,t}\sim k_{y, \min}^{0.73}$, for $k_y\rho_i=0.28$ having a significant amplitude in the $k_y$-spectra of $|\Phi|^2$ and heat flux. Using this scaling $\partial^2\hat{\rm RS}_{k_y}/\partial x^2\sim|\hat{\Phi}_{k_y}|^2\sim k_{y, \min}^{0.73}$ in place of $\sim k_{y, \min}^{1}$ in (\ref{RSscaling}), we get
    \begin{equation}
    {\rm Var}\left(\sum_{k_y}\frac{\partial^2}{\partial x^2}\hat{\rm RS}_{k_y}\right)\simeq\sum_{k_y}{\rm Var}\left(\frac{\partial^2}{\partial x^2}\hat{\rm RS}_{k_y}\right)\\
    \sim \frac{1}{k_{y,\min}}(k_{y,\min}^{0.73})^{2}\sim k_{y,\min}^{0.46},
    \label{RSscalingmod}
    \end{equation}
\emph{i.e.} the standard deviation of $\partial^2{\rm RS}/\partial x^2$ scales as ${\rm SD}={\rm Var}^{0.5}\sim k_{y, \min}^{0.23}$, which is very close to the observed trend of $\langle{\rm SD}(\partial^2{\rm RS}/\partial x^2)\rangle_{x,t}\sim k_{y, \min}^{0.22}$.

%
%----------------------------------------------------------------------
%
\section{Conclusions}\label{Conclusions}
In this paper, we have addressed the self-interaction mechanism and how it affects the drive of zonal flows in ion-scale gyrokinetic turbulence simulations. The basic mechanism of self-interaction is the process by which a microinstability eigenmode non-linearly interacts with itself,
%along the parallel (along magnetic field) direction
generating a Reynolds stress contribution localised at its corresponding MRS (figure~\ref{Lind2RS}). Compared to adiabatic electron simulations, this effect is more prominent in kinetic electron simulations. The origin of this increased self-interaction in kinetic electron simulations is the result of the non-adiabatic passing electron response at MRSs leading to fine slab-like structures at these radial positions, reflected as well by broadened ballooning structures of the eigenmodes (figure~\ref{Ballooning}).
    
These self-interacting contributions from various eigenmodes to Reynolds stress radially align at low order MRSs to generate significant stationary zonal flow shear layers at these positions. However, given that low order MRSs occupy only a tiny radial fraction of core tokamak plasmas, we have focused primarily on how self-interaction affects the drive of zonal flows between the low order MRSs. We have found that self-interaction plays an important role in generating fluctuating zonal flows, in fact {\it throughout} the full radial extent. These fluctuating components of zonal flows are furthermore found to be critical to regulating transport levels.
    
These findings were obtained by studying the self-interaction contributions to Reynolds stress drive from the various microturbulence modes by focusing on their statistical properties. Critical to this approach has been to vary the number of significant toroidal modes participating in our turbulence simulations by performing scans over $k_{y,\min}\rho_i$. 

In these turbulence simulations, using correlation analysis (see figure~\ref{normcorr}), we have in particular shown that the amplitude of the Reynolds stress contributions from self-interaction of each microturbulence mode are essentially random and decorrelated with each other. In the case of kinetic electron simulations, the significant self-interaction mechanism can in this way disrupt the coherent contributions from the alternative zonal flow drive process provided by the modulational instability mechanism, as reflected by the corresponding weak bicoherence  analysis estimates presented in figure~\ref{Energy_bicoh}. In simulations with adiabatic electrons however, for which self-interaction is weak, bicoherence estimates reflect strong resonant 3 wave interactions, characteristic of the modulational instability process being dominant in this case. 
    
Using simple statistical scaling arguments we have then demonstrated how such a decorrelated drive from the self-interaction of various microturbulence modes could in turn lead to a decrease in shearing rate associated to fluctuating zonal flows with decreasing $k_{y,\min}\rho_i$ [see kinetic electron results in figure~\ref{omegaeffvskymin}(c)]. Assuming that the zonal flow shearing mechanism is  the dominant saturation mechanism at play, this in turn would lead to an increase in gyro-Bohm normalised heat and particle flux levels as $k_{y,\min}\rho_i$ decreases.

\begin{figure}
\centering
  \includegraphics[width=0.49\linewidth]{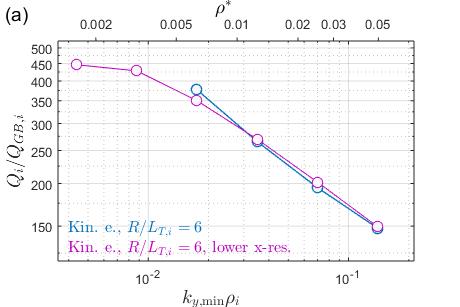}
  \includegraphics[width=0.49\linewidth]{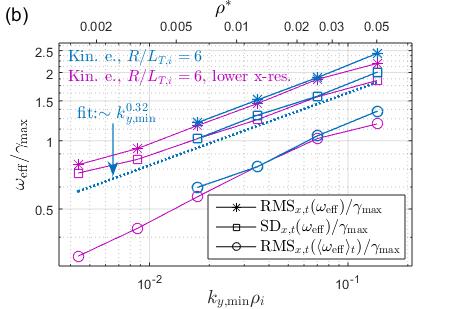}
  \caption{(a) Ion heat flux plotted as a function of $k_{y,\min}\rho_i$. (b) Asterisks, squares and circles represent RMS$_{x,t}(\omega_{\rm eff})$, SD$_{x,t}(\omega_{\rm eff})$ and RMS$_{x}(\langle\omega_{\rm eff}\rangle_t)$ respectively, plotted as a function of $k_{y,\min}\rho_i$. In both sub-plots, blue colour represents the kinetic electron reference simulations whose parameters are given in table~\ref{Parameterset} [the case far from marginal stability ($R/L_{T,i}=6$)]; these are the same plots as those shown with blue colour in figure~\ref{Qivskymin}(a) and figure~\ref{omegaeffvskymin}. Magenta colour represent simulations with a quarter of the resolution in the radial coordinate.}
  \label{Qivskymin_4_dxb4}
\end{figure}

  For the range $k_{y,\min}\rho_i=10^{-2}- 10^{-1}$ considered in section~\ref{Simsetup}, the kinetic electron simulations far from marginal stability [blue circles in figure~\ref{Qivskymin}(a)] indeed show such an increase in ion heat flux, presenting a power law scaling of the form $(k_{y,\min}\rho_i)^{\alpha}$, with $\alpha < 0$. It is important to note that, as one continues to decrease $k_{y,\min}\rho_i$, this scaling ultimately breaks and the simulation results do finally converge; in other words, the true $\rho^*\rightarrow 0$ limit of the flux-tube model can be reached, as already reported in the work by~\citet{Justin2020}. To demonstrate this for the particular case far from marginal stability ($R/L_{T,i}=6$), the original scan was extended to lower values of $k_{y,\min}\rho_i$, but with a quarter of the resolution in the radial direction to make the runs computationally feasible. The corresponding ion heat flux plot is shown in figure~\ref{Qivskymin_4_dxb4}(a), where the power law scaling finally breaks for $k_{y,\min}\rho_i\lesssim 10^{-2}$, and a convergence within $5\%$ is observed at $k_{y,\min}\rho_i\sim 5\times 10^{-3}$. The corresponding estimates of the effective shearing rate, namely RMS$_{x,t}(\omega_{\rm eff})$, SD$_{x,t}(\omega_{\rm eff})$ and RMS$_{x}(\langle\omega_{\rm eff}\rangle_t)$, are plotted in figure~\ref{Qivskymin_4_dxb4}(b). As had been discussed in \S\ref{Simsetup}, the most interesting quantity to us is the shearing rate of fluctuating zonal flows measured by SD$_{x,t}(\omega_{\rm eff})$, which is also seen to deviate from its the power law scaling for values of $k_{y,\min}\rho_i\lesssim 10^{-2}$, hinting towards a near convergence for $k_{y,\min}\rho_i\sim 5\times 10^{-3}$. Furthermore, in figure~\ref{Corrvskymin}, corresponding results of the radial correlation length of turbulent eddies is given, also showing a convergence for $k_{y,\min}\rho_i\sim 5\times 10^{-3}$.

Closer to marginal stability [green line in figure~\ref{Qivskymin}(a)], the gyro-Bohm normalised flux levels in kinetic electrons simulations appear to be already essentially converged for $k_{y,\min}\rho_i\sim 10^{-2}$. The convergence of fluxes observed for this particular case is despite a decrease in the shearing rate associated to zonal flows (seen in RMS$_{x,t}(\omega_{\rm eff})$, SD$_{x,t}(\omega_{\rm eff})$ and RMS$_{x}(\langle\omega_{\rm eff}\rangle_t)$, shown with green lines in figure~\ref{omegaeffvskymin}(a-c)). One possible explanation for this could be that, as the various estimates of normalised shearing rate decrease below $\omega_{\rm eff}/\gamma_{\max}\simeq 1$, the zonal flow saturation mechanism becomes less effective in regulating flux levels. Other saturation mechanisms such as that via damped eigenmodes~\citep{Hatch2011} could then be taking over.

Since $k_{y,\min}\rho_i\sim\rho^*$, one can interpret the decrease in zonal flow shearing rates and the possible associated increase in heat and particle flux levels  in any particular range in $k_{y,\min}\rho_i$ as a finite toroidal system size effect still present in the flux-tube simulations. One may deal in practice with this system size dependence of flux-tube simulations in two ways: 
\begin{enumerate}
    \item[1.] One approach has the intent of correctly resolving this physical finite $\rho^*$ effect, but to be accurate, this requires considering the physical $k_{y,\min} \rho_i$  $=n_{\min}(q_0a/r_0)\rho^*$ value of the tokamak plasma conditions one is studying, which corresponds to the flux-tube covering the {\it full} reference flux-surface in both the poloidal (with $L_z=2\upi$) and toroidal (with $n_{\min}=1$) directions.  In this paper, only this approach has been considered.
    
    This can however be quite challenging in practice, even for flux-tube simulations relevant for a medium-sized tokamak. A scan in $k_{y,\min}$, even when limited to ion scale turbulence, is numerically quite costly, as it involves computations of larger and larger systems along $y$ as $k_{y,\min}$ is decreased. For DIII-D for example, with $\rho^\star\simeq 1/300$, for a full flux-surface located at mid-radius of the plasma ($r_0/a=0.5$), for $q_0=1.4$ considered here, the value of $k_{y,\min}\rho_i=(q_0a/r_0)\rho^* \simeq 1.4\cdot 10^{-2}$ is approximately equal to the value $k_{y,\min}\rho_i =0.0175$ considered in our scan. Large machines such as JET ($\rho^*\simeq 1/600$) and ITER ($\rho^*\simeq 1/1000$) has corresponding values of $k_{y,\min}\rho_i\simeq 4.7\cdot 10^{-3}$ and $2.8\cdot 10^{-3}$ respectively.
    
    Therefore in practice, if the simulation results (including heat and particle fluxes, shearing rate associated to zonal flows, radial correlation length of turbulent eddies etc.) are found to be converged for values of $k_{y,\min}\rho_i$ larger than that corresponding to the machine one is interested in, then one obviously does not need to simulate the full physical flux-surface, as the system size effect of self-interaction has also saturated. However, if no such convergence is observed as $k_{y,\min}\rho_i$ approaches the physical value corresponding to the machine, then one should use that physical $k_{y,\min}\rho_i$ and correctly account for the effect of self-interaction \citep{Justin2020}.

Note that, with this approach of using flux-tube simulations to
physically resolve the particular finite system size effect resulting
from the self-interaction mechanism, the other finite $\rho^*$ effects,
such as profile shearing
  \citep{Waltz1998,Waltz2002}, effect of finite radial extent of the unstable
  region \citep{Ben2010} etc., are missing. Therefore, if one wants to study
how self-interaction competes with other finite $\rho^*$ effects,
either {\it global} simulations should be used, or local simulations
that treat these other finite $\rho^*$ effects explicitly such as \citep{Candy2020}. Note that even in global
simulations, the full flux-surface will have to be modelled to
accurately account for self-interaction, in the sense that a
simulation volume covering only a toroidal wedge, corresponding to considering
$n_{\min}>1$, is in general insufficient.
        
    \item[2.] As a second approach, one can deliberately remove the effects of self-interaction, as suggested by \citet{Beer1995,Faber2018,Justin2020}, by increasing the parallel length of the simulation volume along the magnetic field until convergence is observed. In practice, this is achieved by having the flux-tube undergo multiple poloidal turns before it connects back onto itself, \emph{i.e.} by increasing $L_z={N_{pol}}\cdot 2\pi$ with ${N_{pol}}>1$, instead of $L_z=2\upi$ usually considered. Given that self-interaction is a result of microinstability eigenmodes `biting their tails' after one poloidal turn at corresponding MRSs, this approach weakens the self-interaction mechanism.
    
\end{enumerate}

\begin{acknowledgments}
{\bf Acknowledgemnts:} The authors would like to thank F. Jenko,
T. G\"{o}rler and D. Told for useful discussions pertaining to this
work. This work has been carried out within the framework of the
EUROfusion Consortium and has received funding from the Euratom
research and training programme 2014-2018 and 2019-2020 under grant
agreement no.  633053.  The views and opinions expressed herein do not
necessarily reflect those of the European Commission.  We acknowledge
the CINECA award under the ISCRA initiative, for the availability of
high performance computing resources and support.  Lastly, this work
was supported by a grant from the Swiss National Supercomputing
Centre(CSCS) under project ID s863 and s956.
\end{acknowledgments}
% susie put cite commands here, don't bother with citet etc just yet.

\newpage
\bibliographystyle{jpp}
% Note the spaces between the initials

\bibliography{bibfile_noeprint_cj}

\begin{thebibliography}{40}
\expandafter\ifx\csname natexlab\endcsname\relax\def\natexlab#1{#1}\fi
\def\au#1{#1} \def\ed#1{#1} \def\yr#1{#1}\def\at#1{#1}\def\jt#1{\textit{#1}}
  \def\bt#1{#1}\def\bvol#1{\textbf{#1}} \def\vol#1{#1} \def\pg#1{#1}
  \def\publ#1{#1}\def\arxiv#1{#1}\def\org#1{#1}\def\st#1{\textit{#1}}

\bibitem[Abiteboul(2012)]{AbiteboulPhD}
{\sc \au{Abiteboul, J.}} \yr{2012}  \at{Turbulent and neoclassical toroidal
  momentum transport in tokamak plasmas}. PhD thesis, Aix-Marseille
  Universit{\'e}.

\bibitem[Abiteboul {\em et~al.\/}(2011)Abiteboul, Garbet, Grandgirard, Allfrey,
  Ghendrih, Latu, Sarazin \& Strugarek]{Abiteboul2011}
{\sc \au{Abiteboul, J.}, \au{Garbet, X.}, \au{Grandgirard, V.}, \au{Allfrey,
  S.~J.}, \au{Ghendrih, Ph.}, \au{Latu, G.}, \au{Sarazin, Y.} \& \au{Strugarek,
  A.}} \yr{2011}  \at{Conservation equations and calculation of mean flows in
  gyrokinetics}.  \jt{Physics of Plasmas}  \bvol{18}~(8),  \pg{082503}.

\bibitem[Ball {\em et~al.\/}(2020)Ball, Brunner \& C.J.]{Justin2020}
{\sc \au{Ball, J.}, \au{Brunner, S.} \& \au{C.J., Ajay}} \yr{2020}
  \at{Eliminating turbulent self-interaction through the parallel boundary
  condition in local gyrokinetic simulations}.  \jt{Journal of Plasma Physics}
  \bvol{86}~(2),  \pg{905860207}.

\bibitem[Beer {\em et~al.\/}(1995)Beer, Cowley \& Hammett]{Beer1995}
{\sc \au{Beer, M.~A.}, \au{Cowley, S.~C.} \& \au{Hammett, G.~W.}} \yr{1995}
  \at{Field-aligned coordinates for nonlinear simulations of tokamak
  turbulence}.  \jt{Physics of Plasmas}  \bvol{2}~(7),  \pg{2687--2700}.

\bibitem[Biglari {\em et~al.\/}(1990)Biglari, Diamond \& Terry]{Biglari1990}
{\sc \au{Biglari, H.}, \au{Diamond, P.~H.} \& \au{Terry, P.~W.}} \yr{1990}
  \at{Influence of sheared poloidal rotation on edge turbulence}.  \jt{Physics
  of Fluids B: Plasma Physics}  \bvol{2}~(1),  \pg{1--4}.

\bibitem[Candy {\em et~al.\/}(2020)Candy, Belli \& Staebler]{Candy2020}
{\sc \au{Candy, J}, \au{Belli, E~A} \& \au{Staebler, G}} \yr{2020}
  \at{Spectral treatment of gyrokinetic profile curvature}.  \jt{Plasma Physics
  and Controlled Fusion}  \bvol{62}~(4),  \pg{042001}.

\bibitem[Chen {\em et~al.\/}(2000)Chen, Lin \& White]{Chen2000}
{\sc \au{Chen, Liu}, \au{Lin, Zhihong} \& \au{White, Roscoe}} \yr{2000}
  \at{Excitation of zonal flow by drift waves in toroidal plasmas}.
  \jt{Physics of Plasmas}  \bvol{7}~(8),  \pg{3129--3132}.

\bibitem[Chowdhury {\em et~al.\/}(2008)Chowdhury, Ganesh, Angelino, Vaclavik,
  Villard \& Brunner]{Chowdhury2008}
{\sc \au{Chowdhury, J.}, \au{Ganesh, R.}, \au{Angelino, P.}, \au{Vaclavik, J.},
  \au{Villard, L.} \& \au{Brunner, S.}} \yr{2008}  \at{Role of non-adiabatic
  untrapped electrons in global electrostatic ion temperature gradient driven
  modes in a tokamak}.  \jt{Physics of Plasmas}  \bvol{15}~(7),  \pg{072117}.

\bibitem[Connor {\em et~al.\/}(1978)Connor, Hastie \& Taylor]{Connor1978}
{\sc \au{Connor, J.~W.}, \au{Hastie, R.~J.} \& \au{Taylor, J.~B.}} \yr{1978}
  \at{Shear, periodicity, and plasma ballooning modes}.  \jt{Physical Review
  Letters}  \bvol{40},  \pg{396--399}.

\bibitem[Diamond {\em et~al.\/}(2005)Diamond, Itoh, Itoh \& Hahm]{Diamond2005}
{\sc \au{Diamond, P.~H.}, \au{Itoh, S-I}, \au{Itoh, K.} \& \au{Hahm, T.~S.}}
  \yr{2005}  \at{Zonal flows in plasma - a review}.  \jt{Plasma Physics and
  Controlled Fusion}  \bvol{47}~(12A),  \pg{R35}.

\bibitem[Dimits {\em et~al.\/}(2000)Dimits, Bateman, Beer, Cohen, Dorland,
  Hammett, Kim, Kinsey, Kotschenreuther, Kritz, Lao, Mandrekas, Nevins, Parker,
  Redd, Shumaker, Sydora \& Weiland]{Dimits2000}
{\sc \au{Dimits, A.~M.}, \au{Bateman, G.}, \au{Beer, M.~A.}, \au{Cohen, B.~I.},
  \au{Dorland, W.}, \au{Hammett, G.~W.}, \au{Kim, C.}, \au{Kinsey, J.~E.},
  \au{Kotschenreuther, M.}, \au{Kritz, A.~H.}, \au{Lao, L.~L.}, \au{Mandrekas,
  J.}, \au{Nevins, W.~M.}, \au{Parker, S.~E.}, \au{Redd, A.~J.}, \au{Shumaker,
  D.~E.}, \au{Sydora, R.} \& \au{Weiland, J.}} \yr{2000}  \at{Comparisons and
  physics basis of tokamak transport models and turbulence simulations}.
  \jt{Physics of Plasmas}  \bvol{7}~(3),  \pg{969--983}.

\bibitem[Dominski {\em et~al.\/}(2015)Dominski, Brunner, G{\"o}rler, Jenko,
  Told \& Villard]{Dominski2015}
{\sc \au{Dominski, J.}, \au{Brunner, S.}, \au{G{\"o}rler, T.}, \au{Jenko, F.},
  \au{Told, D.} \& \au{Villard, L.}} \yr{2015}  \at{How non-adiabatic passing
  electron layers of linear microinstabilities affect turbulent transport}.
  \jt{Physics of Plasmas}  \bvol{22}~(6),  \pg{062303}.

\bibitem[Dominski {\em et~al.\/}(2017)Dominski, McMillan, Brunner, Merlo, Tran
  \& Villard]{Dominski2017}
{\sc \au{Dominski, J.}, \au{McMillan, B.~F.}, \au{Brunner, S.}, \au{Merlo, G.},
  \au{Tran, T.-M.} \& \au{Villard, L.}} \yr{2017}  \at{An arbitrary wavelength
  solver for global gyrokinetic simulations. application to the study of fine
  radial structures on microturbulence due to non-adiabatic passing electron
  dynamics}.  \jt{Physics of Plasmas}  \bvol{24}~(2),  \pg{022308}.

\bibitem[Dorland {\em et~al.\/}(2000)Dorland, Jenko, Kotschenreuther \&
  Rogers]{Dorland2000}
{\sc \au{Dorland, W.}, \au{Jenko, F.}, \au{Kotschenreuther, M.} \& \au{Rogers,
  B.~N.}} \yr{2000}  \at{Electron temperature gradient turbulence}.
  \jt{Physical Review Letters}  \bvol{85},  \pg{5579--5582}.

\bibitem[Faber {\em et~al.\/}(2018)Faber, Pueschel, Terry, Hegna \&
  Roman]{Faber2018}
{\sc \au{Faber, B.~J.}, \au{Pueschel, M.~J.}, \au{Terry, P.~W.}, \au{Hegna,
  C.~C.} \& \au{Roman, J.~E.}} \yr{2018}  \at{Stellarator microinstabilities
  and turbulence at low magnetic shear}.  \jt{Journal of Plasma Physics}
  \bvol{84}~(5),  \pg{905840503}.

\bibitem[Falchetto {\em et~al.\/}(2003)Falchetto, Vaclavik \&
  Villard]{Falchetto2003}
{\sc \au{Falchetto, G.~L.}, \au{Vaclavik, J.} \& \au{Villard, L.}} \yr{2003}
  \at{Global-gyrokinetic study of finite {$\beta$} effects on linear
  microinstabilities}.  \jt{Physics of Plasmas}  \bvol{10}~(5),
  \pg{1424--1436}.

\bibitem[Gallagher {\em et~al.\/}(2012)Gallagher, Hnat, Connaughton, Nazarenko
  \& Rowlands]{Gallagher2012}
{\sc \au{Gallagher, S.}, \au{Hnat, B.}, \au{Connaughton, C.}, \au{Nazarenko,
  S.} \& \au{Rowlands, G.}} \yr{2012}  \at{The modulational instability in the
  extended {Hasegawa-Mima} equation with a finite larmor radius}.  \jt{Physics
  of Plasmas}  \bvol{19}~(12),  \pg{122115}.

\bibitem[G{\"o}rler {\em et~al.\/}(2011)G{\"o}rler, Lapillonne, Brunner,
  Dannert, Jenko, Merz \& Told]{GENE2}
{\sc \au{G{\"o}rler, T.}, \au{Lapillonne, X.}, \au{Brunner, S.}, \au{Dannert,
  T.}, \au{Jenko, F.}, \au{Merz, F.} \& \au{Told, D.}} \yr{2011}  \at{The
  global version of the gyrokinetic turbulence code gene}.  \jt{Journal of
  Computational Physics}  \bvol{230}~(18),  \pg{7053 -- 7071}.

\bibitem[Hahm {\em et~al.\/}(1999)Hahm, Beer, Lin, Hammett, Lee \&
  Tang]{Hahm1999}
{\sc \au{Hahm, {T. S.}}, \au{Beer, {M. A.}}, \au{Lin, Z.}, \au{Hammett, {G.
  W.}}, \au{Lee, {W. W.}} \& \au{Tang, {W. M.}}} \yr{1999}  \at{Shearing rate
  of time-dependent {E}$\times ${B} flow}.  \jt{Physics of Plasmas}
  \bvol{6}~(2-3),  \pg{922--926}.

\bibitem[Hallatschek \& Dorland(2005)]{Hallatscheck2005}
{\sc \au{Hallatschek, K.} \& \au{Dorland, W.}} \yr{2005}  \at{Giant electron
  tails and passing electron pinch effects in tokamak-core turbulence}.
  \jt{Physical Review Letters}  \bvol{95},  \pg{055002}.

\bibitem[Hasegawa {\em et~al.\/}(1979)Hasegawa, Maclennan \&
  Kodama]{Hasegawa1979}
{\sc \au{Hasegawa, A.}, \au{Maclennan, C.~G.} \& \au{Kodama, Y.}} \yr{1979}
  \at{Nonlinear behavior and turbulence spectra of drift waves and rossby
  waves}.  \jt{The Physics of Fluids}  \bvol{22}~(11),  \pg{2122--2129}.

\bibitem[Hasegawa \& Mima(1978)]{Hasegawa1978}
{\sc \au{Hasegawa, A.} \& \au{Mima, K.}} \yr{1978}
  \at{Pseudo-three-dimensional turbulence in magnetized nonuniform plasma}.
  \jt{The Physics of Fluids}  \bvol{21}~(1),  \pg{87--92}.

\bibitem[Hatch {\em et~al.\/}(2011)Hatch, Terry, Jenko, Merz \&
  Nevins]{Hatch2011}
{\sc \au{Hatch, D.~R.}, \au{Terry, P.~W.}, \au{Jenko, F.}, \au{Merz, F.} \&
  \au{Nevins, W.~M.}} \yr{2011}  \at{Saturation of gyrokinetic turbulence
  through damped eigenmodes}.  \jt{Physical Review Letters}  \bvol{106},
  \pg{115003}.

\bibitem[Hazeltine \& Newcomb(1990)]{Hazeltine1990}
{\sc \au{Hazeltine, R.~D.} \& \au{Newcomb, W.~A.}} \yr{1990}  \at{Inversion of
  the ballooning transformation}.  \jt{Physics of Fluids B: Plasma Physics}
  \bvol{2}~(1),  \pg{7--10}.

\bibitem[Horton(1999)]{Horton1999}
{\sc \au{Horton, W.}} \yr{1999}  \at{Drift waves and transport}.  \jt{Reviews
  of Modern Physics}  \bvol{71},  \pg{735--778}.

\bibitem[Jenko {\em et~al.\/}(2000)Jenko, Dorland, Kotschenreuther \&
  Rogers]{GENE1}
{\sc \au{Jenko, F.}, \au{Dorland, W.}, \au{Kotschenreuther, M.} \& \au{Rogers,
  B.~N.}} \yr{2000}  \at{Electron temperature gradient driven turbulence}.
  \jt{Physics of Plasmas}  \bvol{7}~(5),  \pg{1904--1910}.

\bibitem[Krommes \& Kim(2000)]{Krommes2000}
{\sc \au{Krommes, John~A.} \& \au{Kim, Chang-Bae}} \yr{2000}  \at{Interactions
  of disparate scales in drift-wave turbulence}.  \jt{Physical Review E}
  \bvol{62},  \pg{8508--8539}.

\bibitem[Lapillonne {\em et~al.\/}(2009)Lapillonne, Brunner, Dannert, Jolliet,
  Marinoni, Villard, Gorler, Jenko \& Merz]{Lapillonne2009}
{\sc \au{Lapillonne, X.}, \au{Brunner, S.}, \au{Dannert, T.}, \au{Jolliet, S.},
  \au{Marinoni, A.}, \au{Villard, L.}, \au{Gorler, T.}, \au{Jenko, F.} \&
  \au{Merz, F.}} \yr{2009}  \at{Clarifications to the limitations of the
  s-alpha equilibrium model for gyrokinetic computations of turbulence}.
  \jt{Physics of Plasmas}  \bvol{16}~(3),  \pg{032308}.

\bibitem[Lin {\em et~al.\/}(1998)Lin, Hahm, Lee, Tang \& White]{Lin1998}
{\sc \au{Lin, Z.}, \au{Hahm, T.~S.}, \au{Lee, W.~W.}, \au{Tang, W.~M.} \&
  \au{White, R.~B.}} \yr{1998}  \at{Turbulent transport reduction by zonal
  flows: Massively parallel simulations}.  \jt{Science}  \bvol{281}~(5384),
  \pg{1835--1837}.

\bibitem[McMillan {\em et~al.\/}(2010)McMillan, Lapillonne, Brunner, Villard,
  Jolliet, Bottino, Goerler \& Jenko]{Ben2010}
{\sc \au{McMillan, B.}, \au{Lapillonne, X.}, \au{Brunner, S.}, \au{Villard,
  L.}, \au{Jolliet, S.}, \au{Bottino, A.}, \au{Goerler, T.} \& \au{Jenko, F.}}
  \yr{2010}  \at{System size effects on gyrokinetic turbulence}.  \jt{Physical
  Review Letters}  \bvol{105},  \pg{155001}.

\bibitem[Merz(2008)]{GENE3}
{\sc \au{Merz, F.}} \yr{2008}  \at{Gyrokinetic simulation of multimode plasma
  turbulence}. PhD thesis, Universit\"at M\"unster.

\bibitem[Parra \& Catto(2009)]{Parra2009}
{\sc \au{Parra, Felix~I} \& \au{Catto, Peter~J}} \yr{2009}  \at{Vorticity and
  intrinsic ambipolarity in turbulent tokamaks}.  \jt{Plasma Physics and
  Controlled Fusion}  \bvol{51}~(9),  \pg{095008}.

\bibitem[Rosenbluth \& Hinton(1998)]{Rosenbluth1998}
{\sc \au{Rosenbluth, M.~N.} \& \au{Hinton, F.~L.}} \yr{1998}  \at{Poloidal flow
  driven by ion-temperature-gradient turbulence in tokamaks}.  \jt{Physical
  Review Letters}  \bvol{80},  \pg{724--727}.

\bibitem[Scott(1998)]{Scott1998}
{\sc \au{Scott, B.}} \yr{1998}  \at{Global consistency for thin flux tube
  treatments of toroidal geometry}.  \jt{Physics of Plasmas}  \bvol{5}~(6),
  \pg{2334--2339}.

\bibitem[Waltz(2005)]{Waltz2005b}
{\sc \au{Waltz, R.~E.}} \yr{2005}  \at{Rho-star scaling and physically
  realistic gyrokinetic simulations of transport in diii-d}.  \jt{Fusion
  Science and Technology}  \bvol{48}~(2),  \pg{1051--1059}.

\bibitem[Waltz {\em et~al.\/}(2006)Waltz, Austin, Burrell \& Candy]{Waltz2006}
{\sc \au{Waltz, R.~E.}, \au{Austin, M.~E.}, \au{Burrell, K.~H.} \& \au{Candy,
  J.}} \yr{2006}  \at{Gyrokinetic simulations of off-axis minimum-q profile
  corrugations}.  \jt{Physics of Plasmas}  \bvol{13}~(5),  \pg{052301}.

\bibitem[Waltz {\em et~al.\/}(2002)Waltz, Candy \& Rosenbluth]{Waltz2002}
{\sc \au{Waltz, R.~E.}, \au{Candy, J.~M.} \& \au{Rosenbluth, M.~N.}} \yr{2002}
  \at{Gyrokinetic turbulence simulation of profile shear stabilization and
  broken gyrobohm scaling}.  \jt{Physics of Plasmas}  \bvol{9}~(5),
  \pg{1938--1946}.

\bibitem[Waltz {\em et~al.\/}(1998)Waltz, Dewar \& Garbet]{Waltz1998}
{\sc \au{Waltz, R.~E.}, \au{Dewar, R.~L.} \& \au{Garbet, X.}} \yr{1998}
  \at{Theory and simulation of rotational shear stabilization of turbulence}.
  \jt{Physics of Plasmas}  \bvol{5}~(5),  \pg{1784--1792}.

\bibitem[Waltz {\em et~al.\/}(1994)Waltz, Kerbel \& Milovich]{Waltz1994}
{\sc \au{Waltz, R.~E.}, \au{Kerbel, G.~D.} \& \au{Milovich, J.}} \yr{1994}
  \at{Toroidal gyro‐landau fluid model turbulence simulations in a nonlinear
  ballooning mode representation with radial modes}.  \jt{Physics of Plasmas}
  \bvol{1}~(7),  \pg{2229--2244}.

\bibitem[Weikl {\em et~al.\/}(2018)Weikl, Peeters, Rath, Seiferling, Buchholz,
  Grosshauser \& Strintzi]{Weikl2018}
{\sc \au{Weikl, A.}, \au{Peeters, A.~G.}, \au{Rath, F.}, \au{Seiferling, F.},
  \au{Buchholz, R.}, \au{Grosshauser, S.~R.} \& \au{Strintzi, D.}} \yr{2018}
  \at{The occurrence of staircases in {ITG} turbulence with kinetic electrons
  and the zonal flow drive through self-interaction}.  \jt{Physics of Plasmas}
  \bvol{25}~(7),  \pg{072305}.

\end{thebibliography}

\end{document}